\documentclass[aip,jcp,reprint,twocolumn,preprintnumbers,amsmath,amssymb,superscriptaddress,nofootinbib,longbibliography]{revtex4-1} 
\pdfoutput=1

\usepackage{graphicx}
\usepackage{amsmath}
\usepackage{amssymb}

\usepackage{etoolbox} 
\usepackage{xspace}
\usepackage{dcolumn}
\usepackage{float}
\usepackage[tight]{subfigure}
\usepackage{verbatim}
\usepackage{enumitem}
\usepackage{units}

\usepackage[usenames,dvipsnames]{xcolor}

\usepackage{fdsymbol} 
\usepackage{bm}
\usepackage[normalem]{ulem}
\usepackage{enumitem}

\renewrobustcmd{\footnote}[1]{ {#1}\xspace}

 \newrobustcmd{\g}{\text{g}}
 \renewrobustcmd{\L}{\text{L}}
 \newrobustcmd{\R}{\text{R}}
 \newrobustcmd{\A}{\text{A}}
 \newrobustcmd{\T}{\text{T}}
 \newrobustcmd{\res}{\text{res}}
\renewrobustcmd{\H}{\mathcal{H}}

\renewcommand{\vec}[1]{\mathbf{#1}}

\newcommand{\Sec}[1]{Sec.~\ref{#1}}
\renewcommand{\sec}[1]{\ref{#1}}
\newcommand{\Eq}[1]{Eq.~(\ref{#1})}
\newcommand{\Eqs}[1]{Eqs.~(\ref{#1})}
\newcommand{\eq}[1]{(\ref{#1})}
\newcommand{\Fig}[1]{Fig.~\ref{#1}}
\newcommand{\Tab}[1]{Table~\ref{#1}}
\newcommand{\fig}[1]{\ref{#1}}
\newcommand{\Ref}[1]{Ref.~\onlinecite{#1}} 
\newcommand{\Refs}[1]{Refs.~\onlinecite{#1}}  

\newrobustcmd{\qd}{resonant\xspace}
\newrobustcmd{\Qd}{Resonant\xspace}
\newrobustcmd{\stm}{off-resonant\xspace}
\newrobustcmd{\Stm}{Off-resonant\xspace}
\newrobustcmd{\set}{SET\xspace}
\newrobustcmd{\Set}{SET\xspace}
\renewrobustcmd{\cot}{COT\xspace}
\newrobustcmd{\Cot}{COT\xspace}
\newrobustcmd{\ecot}{ECOT\xspace}
\newrobustcmd{\Ecot}{ECOT\xspace\xspace}
\newrobustcmd{\icot}{ICOT\xspace}
\newrobustcmd{\Icot}{ICOT\xspace}
\newrobustcmd{\asw}{ASW\xspace}
\newrobustcmd{\Asw}{ASW\xspace}
\newrobustcmd{\coset}{COSET\xspace}
\newrobustcmd{\Coset}{COSET\xspace}

\newrobustcmd{\didv}{d$I$$/$d$V$\xspace}
\newrobustcmd{\didvv}{d$^2I$$/$d$V^2$\xspace}

\begin{document}
\title{
Transport mirages in single-molecule devices
}
\author{R. Gaudenzi}
\affiliation{Kavli Institute of Nanoscience, Delft University of Technology, 2600 GA Delft, The~Netherlands}
\author{M. Misiorny}
\affiliation{Department of Microtechnology and Nanoscience MC2, Chalmers University of Technology, 412 96 G\"{o}teborg, Sweden}
\affiliation{Faculty of Physics, Adam Mickiewicz University, 61-614 Pozna\'{n}, Poland}
\author{E. Burzur{\'i}}
\affiliation{Kavli Institute of Nanoscience, Delft University of Technology, 2600 GA  Delft, The~Netherlands}
\author{M. R. Wegewijs}
\affiliation{Peter Gr{\"u}nberg Institut, Forschungszentrum J{\"u}lich, 52425 J{\"u}lich,  Germany}
\affiliation{JARA-FIT, 52056 Aachen, Germany}
\affiliation{Institute for Theory of Statistical Physics, RWTH Aachen, 52056 Aachen,  Germany}
\author{H. S. J. van der Zant}
\affiliation{Kavli Institute of Nanoscience, Delft University of Technology, 2600 GA  Delft, The~Netherlands}

\begin{abstract}
Molecular systems can exhibit a complex, chemically tailorable inner structure which allows for targeting of specific mechanical, electronic and optical properties. At the single-molecule level, two major complementary ways to explore these properties are molecular quantum-dot structures and scanning probes. This article outlines comprehensive principles of electron-transport spectroscopy relevant to both these approaches
and presents a new, high-resolution experiment on a high-spin single-molecule junction exemplifying these principles.
Such spectroscopy plays a key role in further advancing our understanding of molecular and atomic systems, in particular the relaxation of their spin.
In this joint experimental and theoretical analysis, particular focus is put on the crossover between \emph{resonant} regime [single-electron tunneling (SET)]
and the \emph{off-resonant} regime [inelastic electron (co)tunneling (IETS)]. We show that the interplay of these two processes leads to unexpected \emph{mirages} of resonances not captured by either of the two pictures alone.
Although this turns out to be important in a large fraction of the possible regimes of level positions and bias voltages,
it  has been given little attention in molecular transport studies.
Combined with nonequilibrium IETS -- four-electron pump-probe excitations -- these mirages provide crucial information on the relaxation of spin excitations.
Our encompassing physical picture is supported by a master-equation approach that goes beyond weak coupling.
The present work encourages the development of a broader connection between the fields of molecular quantum-dot and scanning probe spectroscopy.
\end{abstract}

\maketitle
\section{Introduction\label{sec:intro}}
%
\begin{figure}[t]
  \includegraphics[width=0.99\columnwidth]{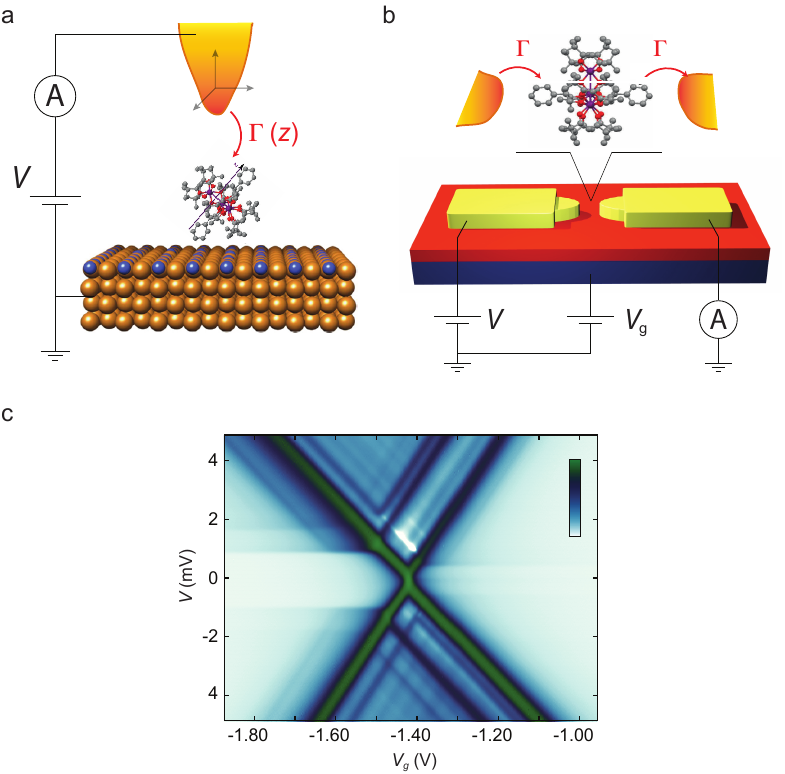}
  \caption{\textbf{High-spin single-molecule junctions.}
    (a) Vertical approach: the metallic tip of a scanning tunneling microscope
    (STM) allows one to scan laterally in real space
    and acquire transport spectra as a function of the bias $V$ at specific molecular sites.
    The  vertical position $z$ controls the tip-molecule coupling $\Gamma(z)$,
    while the molecule-substrate coupling is fixed.
    (b) Planar approach: by embedding
    a bottom-up synthesized magnetic molecule into a solid-state device,
    one can control its energy levels through a gate-voltage $V_\text{g}$.
    This \emph{scanning in energy space} grants access to both regimes of "real" (redox) charging 
    and "virtual" charging (scattering) and their nontrivial crossover.
    (c) Conductance map showing a range of features in the \qd regime (center),
    \stm regimes (far left and right) as well as the crossover regime.
    These are analyzed in detail in \Fig{fig:10} and \Fig{fig:15}.
  }
  \label{fig:1}
\end{figure}
Both the fundamental and applied studies on transport phenomena in electronic devices of molecular
dimensions have bloomed over the past decade\cite{Bartolome_book,Huang_Chem.Soc.Rev.44/2015,Moth-Poulsen_book,Perrin_Chem.Soc.Rev.44/2015}.
An interesting aspect of this development is that it has increasingly hybridized the diverse fields of chemistry, nanofabrication and physics with the primary ambition of accessing properties like high spin and large exchange couplings, vibrational modes, large 
charging energies and long electronic/nuclear spin coherence times, subtle electronic orbital interplay, self-organisation~\cite{Chen_Nat.Nanotech.7/2012,Seeman_Nature421/2003} and chirality \cite{Xie_NanoLett.11/2011, Naaman_RevPhysChem.66/2015}.
This is rendered possible by the higher energy scales of the molecular systems ---a direct consequence of their size--- and 
their complex, chemically tailorable, inner structures which have proven to be effective in addressing, for instance,     
the spin-phonon\cite{Burzuri_NanoLett.14/2014, Franke_J.Phys.:Condens.Matter24/2012}, Shiba\cite{Franke_Science332/2011, Island_arXiv2016} 
and Kondo physics\cite{Frisenda_NanoLett.15/2015}, quantum interference effects\cite{Koole_NanoLett.15/2015} and nuclear spin manipulation\cite{Thiele_Science344/2014}.

In most of the works, in particular those concerning
molecular \emph{spin} systems, two complementary approaches 
have contributed to explore these effects.
On the one hand stands \stm transport spectroscopy, which is the major tool of choice in the scanning-tunneling microscopy (STM) approach to nanoscale spin systems~\cite{Meservey_Phys.Rep.238/1994,Nussinov_Phys.Rev.B68/2003,Heinrich_Science306/2004,Meier_Science320/2008,Chen_Phys.Rev.Lett.101/2008,Wiesendanger_Rev.Mod.Phys.81/2009,Petukhov_Coord.Chem.Rev.253/2009,Kahle_NanoLett.12/2012,Gauyacq_Prog.Surf.Sci.87/2012,Ternes_NewJ.Phys.17/2015,Burgess_NatureComm.6/2015,Heinrich_NanoLett.15/2015},
depicted in \Fig{fig:1}(a).
\Stm is also dominant in the field of mechanically-controlled break junctions (MCBJ)\cite{Boehler_Nanotechnology15/2004,Champagne_NanoLett.5/2005} to study vibrations~\cite{Chae_NanoLett.6/2006,Galperin_J.Phys.:Condens.Matter19/2007,Hihath_NanoLett.8/2008,Huettel_Phys.Rev.Lett.102/2009,Burzuri_NanoLett.14/2014} and, less often, spin effects\cite{Parks_Phys.Rev.Lett.99/2007,Parks_Science328/2010,Frisenda_NanoLett.15/2015}.
On the other hand, \qd transport spectroscopy, originating in the multi-terminal fabrication of quantum dots (QDs, \Fig{fig:1}(b))\cite{Hanson_Rev.Mod.Phys79/2007}, is a well-developed tool applied to a broad range of excitations in nanostructures,~\cite{Jarillo-Herrero_Nature429/2004,Koppens_Science309/2005,Jarillo-Herrero_Phys.Rev.Lett.94/2005,Sapmaz_Phys.Rev.B71/2005,Huettel_Phys.Rev.B72/2005,Graeber_Phys.Rev.B74/2006,Sapmaz_Semicond.Sci.Technol.21/2006,Leturcq_NaturePhys.5/2009,Andergassen_Nanotechnology21/2010,Haupt_Phys.StatusSolidiB250/2013} including spin.\cite{Heersche_Phys.Rev.Lett.96/2006a,Zyazin_NanoLett.10/2010,Burzuri_Phys.Rev.Lett.109/2012,Misiorny_Phys.Rev.B91/2015,Burzuri_J.Phys.:Condens.Matter27/2015,Grose_NatureMater.7/2008,Kogan_Phys.Rev.B67/2003,Craig_Science304/2004,Moriyama_Phys.Rev.Lett.94/2005,Florens_J.Phys.:Condens.Matter23/2011,Grove-Rasmussen_Phys.Rev.Lett.108/2012}
\begin{table*}[tb]
\caption{\label{tab:compare}
\textbf{Nomenclature of \stm and \qd spectroscopy in different communities.}
}
\centering
\renewcommand{\arraystretch}{1.2}
\begin{tabular}{l l l l}
		\hline\hline
		\hspace*{0.5em}\textbf{Regime}\hspace*{0.5em}\rule{0cm}{12pt}&
		\hspace*{0.5em}\textbf{Section}\hspace*{0.5em}&
		\hspace*{0.5em}\textbf{QD community}\hspace*{0.5em}&
		\hspace*{0.5em}\textbf{STM / MCBJ community}\hspace*{0.5em}\\[0.5ex]
		\hline
		\hspace*{0.5em}\Qd \rule{0cm}{12pt}
		& \hspace*{0.5em}\Sec{sec:qd} 
		& \hspace*{0.5em}Single-electron tunneling (SET)
		& \hspace*{0.5em}Resonant tunneling
		\\
		\hspace*{0.5em}
		& \hspace*{0.5em}
		& \hspace*{0.5em}Sequential / incoherent tunneling
		& \hspace*{0.5em}
		\\
		\hspace*{0.5em}\Stm    
		& \hspace*{0.5em}\Sec{sec:stm}
		& \hspace*{0.5em}(In)elastic co-tunneling (COT)
		& \hspace*{0.5em}(In)elastic electron tunneling spectroscopy (EETS/IETS)\hspace*{0.5em}
		\\
		&
		&  \hspace*{0.5em}Coherent tunneling
		&   
		\\
		
		&
		& \hspace*{0.5em}Schrieffer-Wolff (transformation)
		& \hspace*{0.5em}Appelbaum (Hamiltonian)
		\\
		\hspace*{0.5em} 
		& \hspace*{0.5em}\Sec{sec:cot}
		& \hspace*{0.5em} 
		& \hspace*{0.5em}Pump-probe (co)tunneling spectroscopy\hspace*{0.5em}
		\\
		\hspace*{0.5em}Crossover    
		& \hspace*{0.5em}\Sec{sec:crossover}
		& \hspace*{0.5em}Cotunneling-assisted single-electron
		& 
		\\
		
		&
		& \hspace*{0.5em}tunneling (COSET, CAST)
		&
		\\[0.5ex]
		\hline
\end{tabular}
\end{table*}
The key difference  between \qd and \stm approaches is the former's reliance on energy-level control \emph{independent} of the transport bias,
i.e.,
true \emph{gating}
 of the molecular levels\cite{Mason_Science303/2004,Biercuk_NanoLett.5/2005,Hauptmann_NaturePhys.4/2008,Song_Nature462/2009,Jespersen_NaturePhys.7/2011},
which should be distinguished from the capacitive level shift in STM which is \emph{caused} by the bias.
In terms of physical processes, this difference corresponds to
\qd spectroscopy relying on ``real'' charging of the molecule and \stm transport involving only ``virtual'' charging.

In this contribution we discuss a comprehensive picture of transport applicable to a large family of nanoscale objects.  
This is motivated by the experimental spectrum  of a molecular junction depicted in \Fig{fig:1}(c).
Such a conductance map is so full of detail
that it warrants a systematic joint experimental and theoretical study.
In particular, we discuss several effects  which are often overlooked despite their importance to electron transport spectroscopy
and despite existing experimental~\cite{Schleser_Phys.Rev.Lett.94/2005, Huettel_NewJ.Phys.10/2008,Huettel_Phys.Rev.Lett.102/2009} and theoretical works
~\cite{Golovach_Phys.Rev.B69/2004,Aghassi_Appl.Phys.Lett.92/2008,Lueffe_Phys.Rev.B77/2008,Becker_Phys.Rev.B77/2008,Weymann_Phys.Rev.B78/2008,Leijnse_Phys.Rev.B78/2008}.
For instance, it turns out that 
 ``inelastic'' or ``off-resonant'' transport is \emph{not} simply equivalent to the statement that ``resonant processes play no role''.
In fact, we show that generally less than $55 \%$ of the parameter regime of applied voltages that nominally qualified as ``off-resonant'' is actually described by the
widely used inelastic (co)tunneling (\cot or IETS) picture.
Although in many experiments to date this has not been so apparent,
our experimental evidence suggests that this needs consideration.
In theoretical considerations, \qd and \stm transport regimes are often taken as complementary. Our measurements illustrate how this overlooks an important class of relaxation processes.
The breakdown of the \cot picture in the \stm regime
 presents, in fact, new opportunities for studying the
relaxation of molecular spin-excitations which are of importance for
applications.
Interestingly, these resonances are qualitative indicators of a device of
high quality, e.g., for applications involving spin-pumping.
We illustrate \emph{experimentally} the ambiguities that the sole modeling of \stm conductance curves can run into.
For instance, we show that this may lead one to infer quantum states  that do not correspond to real excitations, 
but are simply \emph{mirages} of lower lying excitations,
including their Zeeman splittings.
Although elaborated here for a spin system, our conclusions apply generally, for example to electronic~\cite{Schleser_Phys.Rev.Lett.94/2005} 
and vibrational excitations in nano electro-mechanical systems (NEMS)~\cite{Huettel_NewJ.Phys.10/2008,Huettel_Phys.Rev.Lett.102/2009}.

The outline of the paper is as follows:
In \Sec{sec:theory} we review the physical picture of electron tunneling spectroscopy 
and outline how a given spectrum manifests itself in \qd [\Sec{sec:qd}] and \stm [\Sec{sec:stm}] transport spectra.
In \Sec{sec:crossover} we discuss how these two spectra continuously transform into each other as the energy levels are varied relative to the bias voltage.
With this in hand, we put together a physical picture capturing all discussed effects
which will be subsequently applied to describe the experiment in \Sec{sec:experiment}.

In \Sec{sec:experiment} we follow the reverse path of experimental transport spectroscopy:
We reconstruct the excitation spectrum of a high-spin molecular junction
based on the feature-rich transport spectra as a function of bias voltage, magnetic field, and gate voltage.
Starting from the \stm analysis, we use the boundary conditions imposed by the \qd spectrum 
to resolve a number of ambiguities in the \stm state-assignment.
With the full model in hand we 
highlight two informative transport features:
(i) \emph{nonequilibrium \cot}, i.e., a pump-probe spectroscopy using the electronic analog of Raman transitions
and
(ii) \emph{mirages} of \set resonances that occur well inside the \stm regime.
We conclude with an outlook in \Sec{sec:discussion}.

Since we aim to bring the insights from various communities together,
we summarize in \Tab{tab:compare} the different but equivalent terminology used.
For clarity reasons we set $k_\text{B}=\hbar=e=1$ for the rest of this discussion.

\section{Physical pictures of transport - real vs. virtual charging\label{sec:theory}}

\begin{figure}[t]
  \includegraphics[width=0.99\columnwidth]{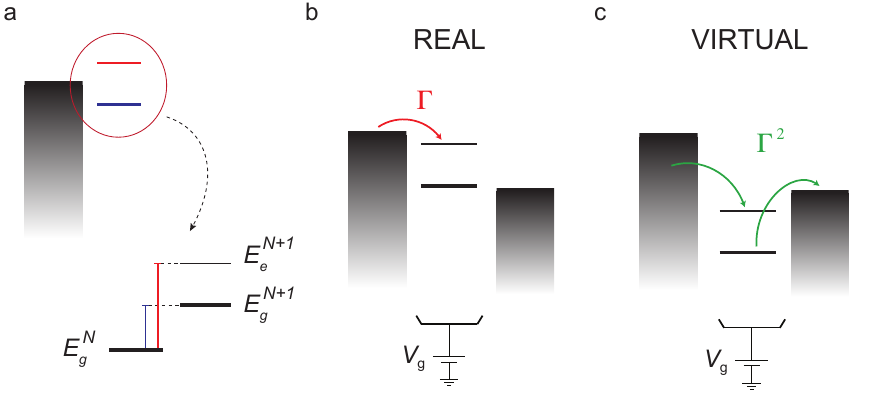}
  \caption{\textbf{Real and virtual transport processes.}
    (a)
    Connection between the molecular energy level and electrochemical potential pictures.  
    Each discrete "level"  in the top panel
    stands for an \emph{electrochemical potential} of the molecule,
    i.e., a \emph{difference} between two energies $E^N_i$
    with different charge $N$ and further quantum numbers denoted by $i$,
    as sketched in the lower panel.
    Due to the capacitive coupling to a gate electrode these energy differences
    can be tuned to be (b) on-resonance and (c) off-resonance with the electrode continuum.
    (b)
    "Real" charging: absorption of an electron,
    reduces the molecule for real, $(N, g) \rightarrow (N + 1, e)$, going from the ground state $g$ 
    for charge $N$ to an excited state $e$ with charge $N + 1$.
    Since this is a one-step process, the rate scales with $\Gamma$, the strength of the
    tunnel coupling.
    (c)
    "Virtual charging": the "scattering" of an electron "off" or "through" the molecule proceeds via any virtual intermediate state,
    for example, starting from the ground state $g$ and ending in a final excited state $e$, $(N + 1, g) \rightarrow (N, g) \rightarrow (N + 1, e)$. 
    The rate of such a two-step process scales as $\Gamma^2$.
    In this case charging is considered only "virtual", as no redox reaction takes place:
    although energy and angular momentum are transferred onto the molecule, the electron number remains fixed to $N + 1$.
  }
  \label{fig:2}
\end{figure}

The two prevalent conceptual approaches to transport through molecular electronic devices
are characterized by the simple physical distinction, sketched in \Fig{fig:2}, between ``real'' charging ---chemical reduction or oxidation---
and ``virtual'' charging ---electrons ``scattering'' between contacts ``through'' a molecular ``bridge''.
Theoretically, the distinction rests on whether the physical processes appear in the leading or next-to-leading order in the 
tunnel coupling strength, $\Gamma$, 
relative to the thermal fluctuation energy $T$.
Experimentally, this translates into distinct applied voltages under which these processes turn on.
These conditions are the primary spectroscopic indicators,
allowing the distinction between ``real'' and ``virtual'' transport processes,
and take precedence over line shape and lifetime broadening.
For reviews on theoretical approaches to molecular transport see \Refs{Thijssen_Phys.StatusSoidiB245/2008,Ferry_book,Cuevas_book,Moth-Poulsen_book}.

\subsection{\Qd transport spectroscopy\label{sec:qd}}

Real charging forms the starting point of what we will call the \qd picture of transport
(see \Tab{tab:compare} for other nomenclature).
Its energy resolution is limited by the Heisenberg lifetime
set by the tunnel coupling $\H^T$
\begin{equation}
	(\H^\T)^2 \propto \Gamma
	,
	\label{eq:lifetime-set}
\end{equation}
allowing for sharp transport spectroscopy of  weakly coupled systems.
This relation has a prominent place in the field of QDs which covers artificial structures as ``artificial atoms'' and 
``artificial molecules'' with redox spectra~\cite{Hanson_Rev.Mod.Phys79/2007} very similar to real atoms~\cite{Tarucha_Phys.Rev.Lett.84/2000} and simple molecules~\cite{Oosterkamp_Nature395/1998,Maksym_J.Phys.:Condens.Matter12/2000,Fujisawa_Nature419/2002,Moriyama_Phys.Rev.Lett.94/2005,Graeber_Phys.Rev.B74/2006}.
Resonant transport also plays a role in STM although its energy resolution 
is often limited by the strong coupling typical of the asymmetric probe-substrate configuration.

Given sufficient weak coupling/energy resolution, 
much is gained when the energy-level dependence of these  transport spectra,
can be mapped out as function of \emph{gate-voltage}.
This dependence allows a detailed model to be extracted
involving just a few  electronic orbitals~\cite{Oosterkamp_Nature395/1998, Begemann_Phys.Rev.B77/2008},
their Coulomb interactions~\cite{Wiel_Rev.Mod.Phys.75/2003}
and their interaction with the most relevant degrees of freedom
(e.g., isotropic~\cite{Grose_NatureMater.7/2008} and
anisotropic spins~\cite{Heersche_Phys.Rev.Lett.96/2006a,Misiorny_Phys.Rev.B79/2009},
quantized vibrations~\cite{Burzuri_NanoLett.14/2014,McCaskey_Phys.Rev.B91/2015},
and nuclear spins~\cite{Vincent_Nature488/2012,Ganzhorn_NatureNanotechn.8/2013,Thiele_Phys.Rev.Lett.111/2013,Thiele_Science344/2014}).
In particular electronic~\cite{Cobden_Phys.Rev.Lett.81/1998,Jarillo-Herrero_Nature429/2004,Biercuk_NanoLett.5/2005,Jarillo-Herrero_Phys.Rev.Lett.94/2005,Sapmaz_Phys.Rev.B71/2005,Graeber_Phys.Rev.B74/2006,Sapmaz_Semicond.Sci.Technol.21/2006,Grove-Rasmussen_Phys.Rev.Lett.108/2012}, spin-orbit~\cite{Jespersen_NaturePhys.7/2011,Jespersen_Phys.Rev.Lett.107/2011} structure as well as electro-mechanical coupling~\cite{Huettel_Phys.Rev.Lett.102/2009,Huettel_NewJ.Phys.10/2008,Leturcq_NaturePhys.5/2009,Weber_NanoLett.15/2015} of CNTs have been very accurately modeled this way.

In molecular electronics transport spectroscopy takes a prominent 
role since imaging of the device is challenging.
By moving to molecular-scale gated structures
one often compromises real-space imaging.
In this paper we highlight the advantages that such structures offer.
Nevertheless,
electrical gates that work simultaneously with a scanning tip~\cite{Piva_Nature435/2005}
or a MCBJ~\cite{Martin10} have been realized,
but with rather low gate coupling.
Notably,  mechanical gating~\cite{Temirov_Nanotech.19/2008,Toher11,Greuling11a,Greuling11b,Greuling13,Wagner_Prog.Surf.Science90/2015}
by lifting a single molecule from the substrate has been demonstrated, resulting in \didv stability diagrams where the role of $V_{\g}$ taken over by the tip-height $z$ in \Fig{fig:1}.
A scanning quantum-dot~\cite{Yoo_Science579/1997} has also been realized
using a single-molecule~\cite{Wagner_Phys.Rev.Lett.115/2015}.

\subsubsection{\Qd excitations - gate dependence\label{sec:qd-gate}}

In the \qd transport regime
one  considers processes of the leading order in the tunnel coupling $\Gamma$,
cf. \Eq{eq:lifetime-set}.
Although most of this is in principle well-known,
we review this approach~\cite{Ferry_book,Bruus_book}
since some of its basic consequences for the \emph{\stm regime} ---discussed below--- are often overlooked.

Typically, analysis of \qd spectra requires a model Hamiltonian $\H$ that involves at most tens of states in the most complex situations~\cite{Koch_Phys.Rev.Lett.94/2005,Koch_Phys.Rev.B74/2006,Grose_NatureMater.7/2008,Leijnse_Phys.Rev.B78/2008,Reckermann_Europhys.Lett.83/2008,Reckermann_Phys.Rev.B79/2009}.
Its energies $E^N_i$ are labeled by the charge number $N$ and a further quantum numbers (orbital, spin, vibrational) collected into an index $i$.
Crucial for the following discussion is the voltage-dependence of this energy spectrum.
We assume it is uniform, i.e., $\propto N$, independent of further quantum numbers $i$.
This can be derived from a capacitive description of the Coulomb interactions between system and electrodes 
referred to as the \emph{constant interaction model}~\cite{Beenakker_Phys.Rev.B44/1991,Wiel_Rev.Mod.Phys.75/2003,Bruus_book,Ferry_book,Thijssen_Phys.StatusSoidiB245/2008}.
In this case,
$E^N_i(V_{\g},V_\text{L},V_{\R}) =
E^N_i
 - N (\alpha_{\g} V_{\g} + \alpha_\text{L} V_\text{L} + \alpha_{\R} V_{\R})$
where $E^N_i$ are constants
and 
$V_\text{L}$ ($V_{\R}$) is the potential applied at source (drain) electrode.
Here, $\alpha_\text{x}=C_x/C$ for $x=\text{L}$, $\text{R}$,$\text{g}$ are capacitive parameters
of which only two are independent since $C:=\sum_x C_x$.
In \Sec{sec:break} we discuss corrections to this ---often good--- assumption\cite{Kaasbjerg_Phys.Rev.B84/2011}.
Unless stated otherwise,
we will set for simplicity $\alpha_{\g}=1$,
i.e., the negative shift of the energy levels equals the gate voltage.
The bias is applied to the electron source, $V_\text{L}=-V$,
and the drain is grounded, $V_{\R}=0$,
giving
$E^N_i(V_{\g},V) = E^N_i -N \alpha_{\g} V_{\g} + N \alpha_\text{L} V$
 and $\mu_\text{L}=\mu_{\R} + V$ with constant $\mu_{\R}$.
Unless stated otherwise, schematics are drawn assuming
$\alpha_\text{L} = 1/2$, corresponding to symmetric and dominant source-drain capacitances $C_\text{L}=C_{\R} \gg C_{\g}$.

The Hamiltonian for the complete transport situation takes the generic form $\H^\text{tot}:= \H + \H^\res+\H^\T$
where $\H^\T$ is a sum of tunneling Hamiltonians that each transfers a single electron across one of the junctions to either metal electrodes.
The electrodes, labeled by $r=$L(left), R(right), are described by $\H^\res$ ---essentially through their densities of states--- and by their electrochemical potentials $\mu^r$ and temperature $T$.
For the present purposes this level of detail suffices, e.g., see \Ref{Leijnse_Phys.Rev.B78/2008} for details. 
For a tunneling process involving such a transfer of precisely one electron,
one of the electrochemical potentials has to fulfill 
\begin{align}
  \mu^r \geq E^{N+1}_{f}-E^{N}_{i}
  \quad
  \text{for}
  \quad
  r=\text{L, R}, 
  \label{eq:set-res}
\end{align}
in order for the electron to be injected into an $N$-electron state $i$, resulting in
the final $N+1$-electron state $f$. Below this threshold the state $(N+1,f)$ is ``unstable'',
and decays
back to $(N,i)$ by expelling the electron back into the electrode.
The rate for the injection process, $W^{N+1,N}_{f,i}$,
is given by familiar ``Golden Rule'' expressions and depends on the difference of both sides of \Eq{eq:set-res} relative to temperature $T$.
When the process ``turns on'' by changing $V$, it gives rise to a peak in the differential conductance, \didv, corresponding to a sharp step in current, of width $T$ and height $\sim \Gamma/T \ll 1$ (in units of $e^2/h$)
since we are assuming weak coupling and high temperature.

If the total system conserves both the spin and its projection along some axis (e.g., the $B$-field axis), the rate involves a selection-rule-governed prefactor.
This prefactor is zero unless the change of the molecular spin and its projection satisfy
\begin{align}
  |\Delta S| =1/2 
  \quad
  \text{and} 
  \quad
  |\Delta M|=\pm 1/2
  \label{eq:rules-set}
  .
\end{align}
These conditions reflect the fact that only a single electron is available for transferring spin to the molecule.

Incidentally, we note that this picture is very useful even beyond the weak couplings and high temperatures assumed here.
Close to the resonance defined by condition \eq{eq:set-res}
the transport still shows a peak
which is, however, modified by higher-order corrections.
The width of the current step becomes broadened $\propto \Gamma$,
giving a conductance peak $\sim 1$ in units of $e^2/h$.
Its energy position may shift on the order of $\Gamma$.

\begin{figure}[t]
  \includegraphics[width=0.99\columnwidth]{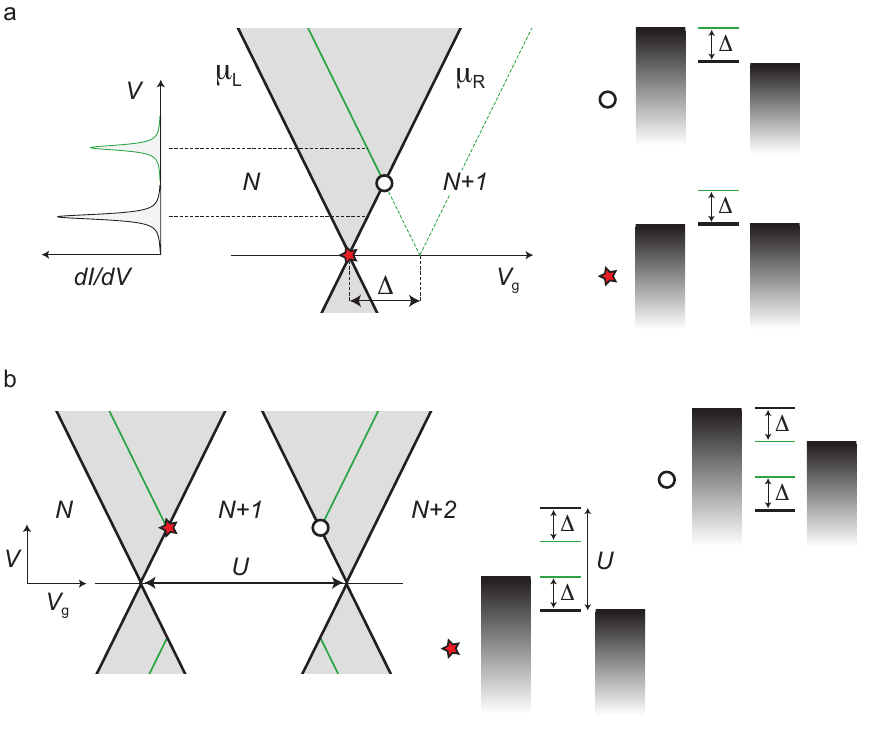}
  \caption{\textbf{Resonant regime: main features}
    (a) Current flows in the bias window set by \Eq{eq:set} (shaded)
    for two charge states $N$ and $N+1$.
    The boundary lines (bold),
    where 
    $\mu_r=E^{N+1}_g-E^{N}_g$ for $r=\L, \R$,
    have slopes $-\alpha_{\g}/(1-\alpha_{\R})$ 
    and $\alpha_{\g}/\alpha_{\R}$, respectively,
    allowing the capacitive parameters to be determined.
    The green lines, offset horizontally by $\Delta =E_e^{N+1}-E_g^{N+1}$, indicate the window of accessibility of the excited state $E_e^{N+1}$
    and are defined by $\mu_r=\Delta + E^{N+1}_g-E^{N}_g$.
    (b) 
    Similar to figure (a),
    for three charge states $N$, $N+1$ and $N+2$.
    This adds a copy of the bias window of (a) that is horizontally offset by the energy $U$ [\Eq{eq:U}] with boundaries
    $\mu_r=E^{N+2}_g-E^{N+1}_g$ for $r=\L,\R$.
    The excitation lines on the right (green) are mirrored horizontally,
    $\mu_r= - \Delta + E^{N+2}_g-E^{N+1}_g$ for $r=\L, \R$,
    since  electron processes relative to $N+1$ have become hole processes.
  }
  \label{fig:set}
\end{figure}

It is now clear in which regime of applied voltages the above picture applies.
In \Fig{fig:set} this is sketched in the plane of applied bias ($V$) and gate voltage ($V_{\g}$).
Here, we call such a (schematic) \didv intensity plot ---also known as ``stability diagram'' or ``Coulomb-diamond''--- a \emph{transport spectrum}.
The indicated vertical line cuts through this diagram correspond to \didv traces measured in STM or MCBJ experiments.
Applied to the ground states of subsequent charge states ---labeled by $g$---, \Eq{eq:set-res} gives
the two inequalities
\begin{align}
  \mu^L \geq E^{N+1}_{g}-E^{N}_{g} \geq \mu^R
  .
  \label{eq:set}
\end{align}
These define the shaded bias window in \Fig{fig:set}(a),
delimited by the ``cross''.
Here,
 a single electron entering from the left can exit to the right,
resulting in a net directed current.

It is now tempting to naively define the \emph{\stm regime} as the complement of the grey \qd regime in \Fig{fig:set}(a),
i.e., by moving across its boundaries by more than $T$ or $\Gamma$.
A key point of our paper is that
this simple rationale is not correct
already for a small finite bias matching some excitation at energy $\Delta$,  
indicated by green lines in \mbox{\Fig{fig:set}(a)}.
Only in the linear-response regime\cite{Beenakker_Phys.Rev.B44/1991} around $\mu=\mu_{\L} =\mu_{\R}$ 
the \stm regime
can be defined as the complement of the resonant regime:
\begin{align}
  | E^{N+1}_{g}-E^{N}_{g} - \mu |  \gg \text{max $\{$} \Gamma, T \}
  \label{eq:naive}
  .
\end{align}

In subsequent charge states
 analogous considerations apply:
transitions between charge states $N+1$ and $N+2$
give rise to a shifted "copy" of the bias window as shown in \Fig{fig:set}(b).
The shift ---experimentally directly accessible--- is denoted by:
\begin{align}
  U :=  (E^{N+2}_{g}-E^{N+1}_{g}) - (E^{N+1}_{g}-E^{N}_{g}) 
  .
  \label{eq:U}
\end{align}
This includes the charging energy of the molecule,
but also the magnitude of
orbital energy differences
and the magnetic field.\footnote{For example, for a single orbital level with charging energy $u > 0$ and magnetic field $B$ one finds $U=u+|B| > u$ due to the opposite spin-filling enforced by the Pauli principle.}

\subsubsection{Stationary state and \qd transport current\label{sec:theory-qd}}

The above rules are substantiated
by a simple master equation 
for the stationary-state occupations $P^{N}_i$ of the states with energy $E^N_i$ 
that can be derived from the outlined model, see, e.g., \Ref{Leijnse_Phys.Rev.B78/2008}.
This approach is used in \Sec{sec:qd-exp} to model part of our experiment.
For the $N\leftrightarrow N+1$ resonance regime the stationary-state equation reads (for notational simplicity we here set $N=0$)
\begin{align}
  \frac{\text{d}}{\text{d}t}
  \begin{bmatrix}
    \vec{P}^{0} \\
    \vec{P}^{1} 
  \end{bmatrix}
   = 0 =
   \begin{bmatrix}
    \vec{W}^{0,0} & \vec{W}^{0,1} \\
    \vec{W}^{1,0} & \vec{W}^{1,1} 
  \end{bmatrix}
  \begin{bmatrix}
    \vec{P}^{0} \\
    \vec{P}^{1} 
  \end{bmatrix}
  \label{eq:set-master}
  .
\end{align}
Here, $\vec{W}^{1,0}$ is the matrix of transition rates $W^{1,0}_{f,i}$ between states $(0,i)$ and $(1,f)$,
and 
analogously for $\vec{W}^{0,1}$.
For example, one of the equations,
\begin{align}
  \frac{\text{d}}{\text{d}t}P^1_f 
  =
  \sum_i W^{1,0}_{f,i} P^0_i
  +W^{1,1}_{f,f} P^1_f
  ,
\end{align}
describes the balance between
the gain in occupation probability due to all transitions $(0,i) \to (1,f)$,
and the leakage $-W^{1,1}_{f,f}$ from the state $(1,f)$.
The entries of the diagonal matrices $\vec{W}^{0,0}$ and $\vec{W}^{1,1}$ have negative values
$W^{0,0}_{f,f}= - \sum_i W^{1,0}_{i,f}$
and
$W^{1,1}_{f,f}= - \sum_i W^{0,1}_{i,f}$,
respectively, such that probability normalization
$\sum_i P^{0}_i+\sum_j P^{1}_j=1$ is preserved in \Eq{eq:set-master}.
In the leading order in $\Gamma$,
the rate matrix has separate contributions from 
the left ($r=\text{L}$) and right ($r=\text{R}$) electrode:
$\vec{W}=\vec{W}^{\L}+\vec{W}^{\R}$.
These allow the stationary current to be computed by counting the electrons transferred by tunnel processes 
through the $r$-th junction,
\begin{align}
  I^r 
  = 
  \sideset{}{'}\sum_{N_f,N_i}
  \sum_{f,i}
  \big(N_f - N_i\big) 
  \times 
  [W^r]_{f,i}^{N_f,N_i} P^{N_i}_{i}
  , 
  \label{eq:set-current}
\end{align}
where stationarity guarantees $I^{\L} = - I^{\R}$. 
We note that, because we are considering only single-electron tunneling processes (first order in $\Gamma$), the primed sum is constrained to $N_f=N_i \pm 1$ by charge conservation.

\subsection{\Stm transport spectroscopy\label{sec:stm}}

We now take the opposite point of view and consider transport entirely due to  ``virtual charging'' or ``scattering through'' the molecule.
The resulting \stm transport spectroscopy, alternatively called cotunneling (\cot) or IETS spectroscopy,
dates back to Lambe and Jacklevic~\cite{Lambe_Phys.Rev.165/1968}.
The discussion of the precise conditions under which the \stm picture applies is postponed to \Sec{sec:crossover}.
Throughout we will denote by the label \cot ---unless stated otherwise---  \emph{inelastic} cotunneling.

The attractive feature  of \stm relative to \qd spectroscopy
is the higher energy resolution as we explain below [\Eq{eq:lifetime-cot} ff.].
Exploiting this in combination with the STM's imaging capability has allowed chemical identification~\cite{Khajetoorians_NaturePhys.8/2012,Kahle_NanoLett.12/2012,Brede_Phys.Rev.B86/2012,Khajetoorians_Science339/2013,Bryant_Phys.Rev.Lett.111/2013,Rau_Science344/2014,Hapala_Phys.Rev.B90/2014,Wagner_NatureCommun.5/2014,Heinrich_NanoLett.15/2015,Burgess_NatureComm.6/2015,Wagner_Phys.Rev.Lett.115/2015,Bazarnik_NanoLett.16/2016}.
This in turn has enabled atomistic modeling of the junction using \textit{ab-initio} calculations~\cite{Hirjibehedin_Science312/2006,Hirjibehedin_Science317/2007,Serrate_NatureNanotech.5/2010,Baumann_Phys.Rev.Lett.115/2015}, also including strong interaction effects~\cite{Greuling_Phys.Rev.B84/2011, Temirov_Phys.Rev.Lett.105/2010,Esat_Phys.Rev.B91/2015},
giving a detailed picture of
transport on the atomic scale~\cite{Atodiresei_Phys.Rev.Lett.102/2009,Loth_NaturePhys.6/2010,Khajetoorians_Phys.Rev.Lett.106/2011,Yan_NanoLett.15/2015,Yan_NatureNanotechnol.10/2015,Khajetoorians_NatureCommun.7/2016,Steinbrecher_NatureCommun.7/2016}.

In recent years, \stm spectroscopy has been also intensively applied to spin systems~\cite{Heinrich_Science306/2004,Otte_NaturePhys.4/2008,Loth_NewJ.Phys.12/2010,Kahle_NanoLett.12/2012,STM_J.Phys.:Condens.Matter26/2014,Ternes_NewJ.Phys.17/2015}
in more symmetric~\cite{Repp_Phys.Rev.Lett94/2005} STM configurations.
However, it is sometimes not realized that the same \stm spectroscopy
also applies to gated molecular junction, and more generally to QDs~\cite{DeFranceschi_Phys.Rev.Lett.86/2001,Zumbuehl04,Katsaros10,Jespersen11}.
In fact, motivated by the enhanced energy resolution,
spectroscopy of discrete
spin-states was introduced in gate-controlled semiconductor QDs~\cite{DeFranceschi_Phys.Rev.Lett.86/2001,Wegewijs01,Hartmann03}
before it was introduced in STM as ``spin-flip'' spectroscopy~\cite{Heinrich_Science306/2004},
see also~\cite{Averin_Phys.Rev.Lett.65/1990,Grabert_book}.
\cot spectroscopy is also used to study molecular properties other than spin, e.g., vibrational states~\cite{Stipe_Science280/1998,Yu04inel,Youngsang_Phys.Rev.Lett.106/2011,Buerkle_Phys.StatusSolidiB250/2013,Boehler_Phys.Rev.B76/2007,Herz_Appl.Phys.Lett.108/2016}.

\begin{figure}[t]
\includegraphics[width=0.99\columnwidth]{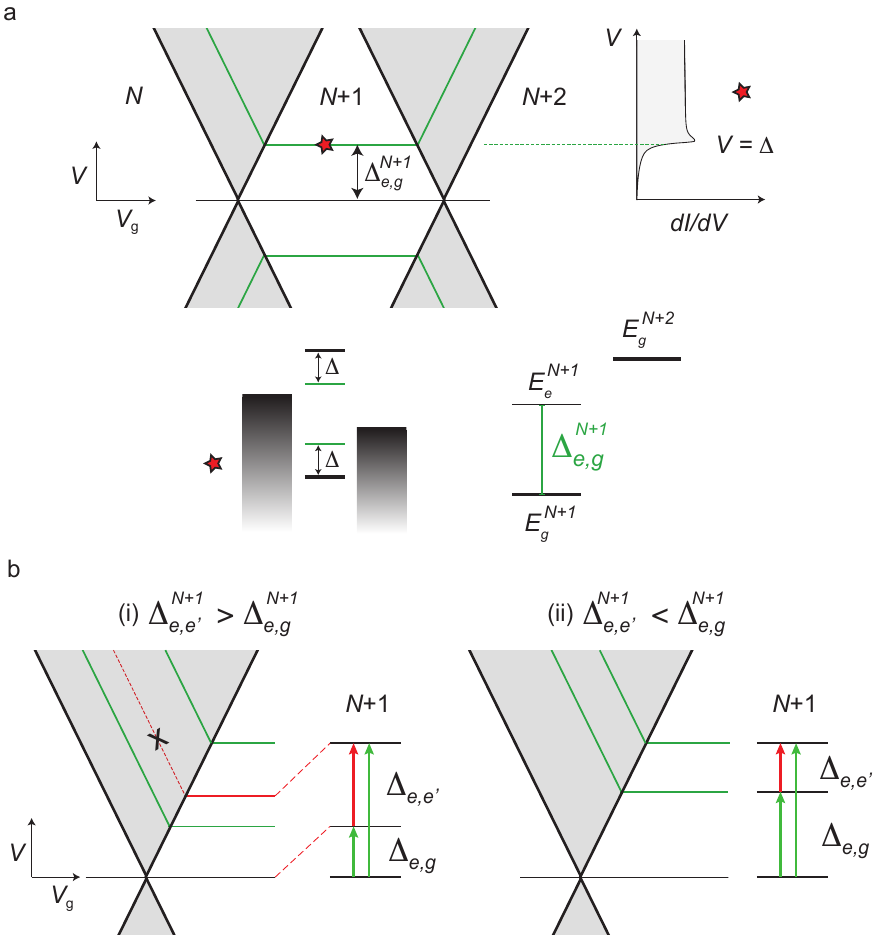}
 \caption{\textbf{\cot transport conditions.}
(a) Same as \Fig{fig:set}(b), now indicating
the "equilibrium" resonance 
(green horizontal line)
in which the excitation $(N+1,e)$ is reached from the ground state $(N+1,g)$ by \cot.
This horizontal line  always connects to the \set resonances for
$(N,g) \to (N+1,e)$ and $(N+2,g) \to (N+1,e)$
(green tilted lines).
(b)~"Nonequilibrium" \cot resonances corresponding to the transition $(N+1,e) \to (N+1,e' )$. 
Case (i) and (ii) are discussed in the text.
Note that there is never a corresponding \set resonance (crossed-out red dashed line) connecting to a non-equilibrium excitation 
as is the case for ``equilibrium'' \cot lines in (a).
}
 \label{fig:cot}
\end{figure}

\subsubsection{\Stm excitations -- no gate dependence
\label{sec:cot}}

In the \stm picture, one considers transport due to next-to-leading order processes,
i.e. of order $\Gamma^2$ in the tunnel rates.
This involves elastic (inelastic) processes involving two electrons from the electrodes
and a zero (net) energy transfer of energy.
When the maximal energy supplied by the electrons ---one electron coming in from, say, $r=\text{L}$ at high energy $\mu_{\L}$, and
the other outgoing to $r=\text{R}$ at low energy $\mu_{\R}$---
exceeds a discrete energy difference of the molecule,
\begin{align}
	\mu_{\L} -\mu_{\R} = V \geq \Delta^{N+1}_{f,i} :=  E^{N+1}_{f}-E^{N+1}_{i}
	,
	\label{eq:cot-res}
\end{align}
transport may be altered with $V$.
Importantly, on the right hand side, all $V$ and $V_{\g}$ dependences of the energies  cancel out [cf. \Sec{sec:qd-gate}]
since we  assumed that the applied voltages \emph{uniformly} shift the excitation spectrum for fixed charge.\cite{Kaasbjerg_Phys.Rev.B84/2011,Perrin_NatureNanotechn.8/2013}

The occurrence of such a process depends on whether the initial state $i$ is occupied
or not by another already active process. It thus depends on whether we are in the ``equilibrium'' or ``nonequilibrium'' regime, both of which are accessible in our experiment in \Sec{sec:experiment}.
The spectroscopy rules require the following separate discussion.

\paragraph{``Equilibrium'' inelastic \cot---}

Already in the linear transport regime,
$V \lesssim T,\Gamma$\footnote{(assuming no excitations lie below  $T$ and $\Gamma$),}
there is scattering through the molecule in a fixed stable charge state in the form of \emph{elastic} \cot~\cite{Averin_Phys.Rev.Lett.65/1990,Grabert_book,Koenig_Phys.Rev.Lett.78/1997}, see \Tab{tab:compare} for the varied nomenclature. This 
gives rise to
a small current scaling $\propto \Gamma^2$.
With increasing bias $V$, this mechanism yields a nonlinear background current
which is, however, featureless.

When the voltage provides enough energy to reach the lowest excitation $e$ of the $N+1$-electron ground state $g$,
the transition 
$(N+1,g) \to (N+1,e)$ is enabled, cf. \Fig{fig:2}(c). This occurs when the \emph{gate-voltage independent} 
criterion set by \Eq{eq:cot-res} with $i=g$ and $f=e$ is satisfied:
\begin{align}
  V \geq \Delta_{e,g}^{N+1}
  \label{eq:cot-res2}
  .
\end{align}
The above energy condition
is the tell-tale sign of an \stm process:
as sketched in \Fig{fig:cot}(a), this allows for  a clear-cut distinction from \qd processes  with a gate dependent energy condition \eq{eq:set-res}.
Importantly, such a \cot feature always connects to the gate-dependent \set resonance corresponding to excitation $\Delta^{N+1}_{e,g}$.
As in the \qd regime, we stress that criterion \eq{eq:cot-res2} uses
the peak position in the ($V_{\g},V)$ plane as a primary indicator. 
The line shape along a vertical cut in the figure, as measured in STM, may be less clear.
In theoretical modeling the line shape is also not a unique indicator.
The line shape is a good secondary indicator of the nature of a process.

\paragraph{``Nonequilibrium'' inelastic \cot:
electronic pump-probe spectroscopy.---}
The above ``equilibrium'' picture of \stm transport
has been successfully applied in many instances.
However, as the first excited state 
$(N+1,e)$
is accessed, the rules of the game change. If the relaxation induced by sources other than transport is weak enough~\cite{Schmaus_Phys.Rev.B79/2009}, 
the occupation of the excited states can become non-negligible. 
In such a case, as illustrated in \Fig{fig:cot}(b), a secondary inelastic \cot process from the excited state $e$ to an even higher excited state $e'$ should be considered. 
Such secondary processes, with the generic condition: 
\begin{align}
  V \geq \Delta^{N+1}_{e'e} = \Delta_{e'g}^{N+1} - \Delta_{eg}^{N+1}
  \label{eq:cot-res-neq}
  ,
\end{align}
indicate a device with an intrinsic relaxation rate  small compared to \cot rates $\propto\Gamma^2$.
As discussed in \Fig{fig:cot}(b),
such excitations \emph{never} connect to a corresponding \set excitation in the transport spectrum.
At this point,  two cases have to be considered, both of which are relevant to the our experiment in ~\Sec{sec:pump-probe}.

(i) If 
$\Delta_{e'e}^{N+1} > \Delta_{eg}^{N+1}$
---i.e., the gaps in the energy spectrum grow with energy---
an extra \emph{"nonequilibrium" inelastic \cot} resonance
at bias 
$V=\Delta_{e'g}^{N+1} - \Delta_{eg}^{N+1}$ appears, as illustrated in panel (i) of \Fig{fig:cot}(b).
This extra resonance is very useful since it provides a further consistency check on the  excitations 
$\Delta_{eg}^{N+1}$ and $\Delta_{e'g}^{N+1} $ observed independently in the \set.\footnote{(If the \set transition to $e'$ is not allowed by a selection rule, the secondary \cot resonance may be the only evidence of this state.)}
Clearly, the intensity of such secondary "nonequilibrium" \cot resonances is generally expected to be lower than the primary ones that start from the ground state.
In \Sec{sec:pump-probe} we will experimentally control this sequential \cot ``electronic pump-probe'' excitations 
by tuning a magnetic field.

(ii) In the opposite case, 
$\Delta_{e'e}^{N+1} < \Delta_{eg}^{N+1}$,
no \emph{extra} \cot excitation related to $e'$ appears:
there is no change in the current at the lower
voltage 
$\Delta_{e'e}^{N+1}$
because the initial state $(N+1,e)$ only becomes occupied at the \emph{higher} voltage 
$\Delta_{eg}^{N+1}$.
This is illustrated in panel (ii) of \Fig{fig:cot}(b).
Examples of both these cases  occur in the \stm spectra of molecular magnets due to the interesting interplay of their easy-axis and transverse anisotropy, see the supplement of \Ref{Zyazin_NanoLett.10/2010}.

\subsubsection{Stationary state and \stm transport current
\label{sec:theory-stm}}

Similar to the \qd case,
the  conditions \eq{eq:cot-res}-\eq{eq:cot-res-neq} are incorporated in a simple stationary master equation for \stm transport
 whose derivation we discuss further below.
In particular, the occupation probabilities $\vec{P}^{N+1}$ in the stationary transport state are determined by (as previously, we put $N=0$)
\begin{align}
  \frac{\text{d}}{\text{d}t}
    \vec{P}^{1} 
   = 0 =
     \vec{W}^{1,1} 
    \vec{P}^{1} 
  \label{eq:cot}
  .
\end{align}
Here, $\vec{W}^{1,1}$ is a matrix  of rates $W^{1,1}_{f,i}$ for transitions
between states $i \to f$.
Since in the \stm regime charging is only ``virtual'', these transitions now occur for a fixed charge state.
The matrix takes the form
$\vec{W}^{1,1} = \sum_{r r'} \vec{W}^{1,1; r,r'}$,
including rate matrices $\vec{W}^{1,1; r,r'}$ for back-scattering from the molecule (to the same electrode, $r=r'$) and scattering through it (between electrodes $r \neq r'$).
The current is obtained by counting 
the net number of electrons transferred from one electrode to the other:
\begin{align}
  I^{\L \rightarrow \R } =
  \sum_{f,i} ( W^{1,1; \R, \L}_{f,i} -  W^{1,1; \L, \R}_{f,i} ) P^1_i
  .
  \label{eq:stm-current}
\end{align}
The inclusion into this picture of the above discussed ``non-equilibrium'' \cot effects depends whether one solves the master equation \eq{eq:cot} or not.
To obtain the simpler description of ``equilibrium'' inelastic \cot [case (i) above] one can insert \emph{by hand} equilibrium populations $P^1_i = e^{-E^1_i/T}/\mathcal{Z}^1$ directly into \Eq{eq:stm-current}.
Solving, instead, \Eq{eq:cot} without further assumptions gives
the ``nonequilibrium'' inelastic \cot case~\cite{Paaske_NaturePhys.2/2006,Ternes_NewJ.Phys.17/2015}
discussed above [case (ii)]. 
In practice, these two extreme limits ---both computable without explicit consideration of intrinsic relaxation--- are always useful to compare
since any more detailed modeling of the intrinsic relaxation will lie somewhere in between'.

The electron tunneling rates in \Eq{eq:cot} are made up entirely of contributions of order $\Gamma^2$.
There are two common ways of computing these rates, and  we now present
the underlying physics relevant for the discussion in \Sec{sec:crossover}.

\paragraph{Appelbaum-Schrieffer-Wolff Hamiltonian.---}

A conceptual connection between the  \stm ``virtual'' charging picture 
 and the \qd picture of ``real'' charging in \Sec{sec:qd}
 emerges naturally when applying the unitary transformation~\cite{Wagner_book}
due to Appelbaum~\cite{Appelbaum_Phys.Rev.Lett.17/1966,Appelbaum_Phys.Rev.154/1967}, Schrieffer and Wolff~\cite{Schrieffer_Phys.Rev.149/1966,Schrieffer_J.Appl.Phys.38/1967,Bravyi_Ann.Phys.326/2011} (\asw)
to the transport \emph{Hamiltonian} $\H^\text{tot}$ [cf. \Sec{sec:qd}].
The effective \asw model obtained in this way 
allows one to easily see the key features of the \stm spectroscopy.

In this approach,
the one-electron tunneling processes described by the Hamiltonian 
$\H^T$, are transformed away and the charge state is fixed by hand to a definite integer.
With that, also all the gate-voltage dependence of resonance \emph{positions}
[\Eq{eq:cot-res2} ff.]
drops out.
This new \asw model is obtained by applying a specially chosen unitary transformation $\mathcal{U}$ to the original Hamiltonian such that:
\begin{align}
  \H^\text{tot} & \to \mathcal{U}  (\H+\H^{\R}+\H^\T) \, \mathcal{U}^\dag  \notag
  \\
                & \approx \H+\H^{\R}+\H^\A + \text{O}(\Gamma^4). 
\end{align}
The term $\H^T \propto \Gamma$ is effectively replaced by $\H^{\A}$,
which involves only $\Gamma^2$ processes and represents exclusively scattering of electrons
``off'' and ``through'' the molecule.
In many cases of interest \cite{Ternes_NewJ.Phys.17/2015} this coupling $\H^{\A}$ contains
terms describing the potential (scalar) and exchange (spin-spin) scattering of electrons and holes with amplitudes $J$ and $K$, respectively:
\begin{align}
  \H^\A = \sum_{r,r'} \left( J_{rr'} \vec{S} \cdot \vec{s}_{rr'} + K_{rr'} N \,  n_{rr'} \right)
  .
  \label{eq:HA}
\end{align}
In the equation above, the
operators $\vec{s}_{rr'}$ ($ n_{rr'}$) describe  spin-(in)dependent intra- [$r=r'$] and inter-electrode [$r \neq r'$] scattering of electrons, see \Ref{Bruus_book} for details.
This scattering is coupled to the molecule through its charge \text{($N$, constant)} and spin ($\vec{S}$).

\emph{Selection rules.}
The \asw coupling $\H^\A$ has selection rules that differ from the original single-electron tunnel coupling $\H^\T$:
\begin{align}
  \Delta S=0,\pm1 \quad \text{ and } \quad \Delta M=0,\pm 1
  .
  \label{eq:rules-cot}
\end{align}
These reflect 
that the two electrons involved in the scattering process
have integer spin 0 or 1 available for exchange with the molecule.
We will apply this in \Sec{sec:spectro}.
This is illustrated by the example model \eq{eq:HA} where 
the spin-operator $\vec{S}$ has matrix elements that obey \Eq{eq:rules-cot}.

\emph{Lifetime.}
After transforming
to this new effective picture, scattering becomes the leading order transport mechanism.
The ``Golden Rule'' approach can be then applied analogously to the case of of the \qd regime,
but now with respect to the \asw scattering $\H^\A$.
In this way \Eq{eq:cot} is obtained together with an expression for the corresponding rate matrix $\vec{W}^{1,1}$.
The \didv given by \Eq{eq:stm-current} shows gate-voltage-independent \emph{steps} at energies set by \Eq{eq:cot-res}. 

Although at high temperatures these steps get thermally broadened~\cite{Lambe_Phys.Rev.165/1968}, 
at low enough $T$ their broadening is smaller than that of the \set peaks.
While calculation of this lineshape requires higher-order contributions to $\vec{W}^{1,1}$,
the relevant energy scale (inverse lifetime) is given by the magnitude of the ``Golden Rule'' rates \emph{for the effective coupling $\H^\A$} scaling as
\begin{align}
	(\H^\A)^2  \propto \Gamma^{2}
     \label{eq:lifetime-cot}
    .
\end{align}
This results in a
much larger lifetime compared to the one from \set [cf. \Eq{eq:lifetime-set}]
due to the role of the interactions on the molecule suppressing charge fluctuations.
The smaller intrinsic broadening is a key advantage of \cot vs. \set spectroscopy~\cite{DeFranceschi_Phys.Rev.Lett.86/2001}.

\emph{Line shape.}
Due to nonequilibrium effects
---i.e., the voltage-dependence of the  occupations obtained by solving \Eq{eq:cot}---
a small peak can develop on top of the \cot step~\cite{Wegewijs01,Hartmann03,Paaske_NaturePhys.2/2006,Sothmann10b}.
Moreover, processes beyond the leading-order in $\H^\A$,
which is all the \cot approach accounts for,
can have a similar effect.
These
turn the \stm tunneling step into a \didv peak
and are in use for more precise modeling of experiments~\cite{Hurley11,Hurley12,Ternes_NewJ.Phys.17/2015}.
Spin-polarization~\cite{Schweflinghaus14} and spin-orbit effects~\cite{Katsaros10,Hurley12}, however, also affect the peak shape and asymmetry.

At low temperatures and sufficiently strong coupling
a nonequilibrium Kondo effect develops
which has been studied in great detail~\cite{Rosch_Phys.Rev.Lett.90/2003,Paaske_Phys.Rev.B69/2004,Paaske_NaturePhys.2/2006,Schoeller_Phys.Rev.B80/2009,Saptsov_Phys.Rev.B86/2012}.
These works show that the peak amplitude is then enhanced
nonperturbatively in the tunnel coupling,
in particular for low lying excitations.
This requires nonequilibrium renormalization group methods beyond the present scope
and we refer to various reviews~\cite{Aleiner_Phys.Rep.358/2002,Pustilnik_J.Phys.:Condens.Matter16/2004,Glazman_LesHouches_2005,Schoeller_Euro.Phys.J.SpecialTopics168/2009,Fritsch_Ann.Phys.324/2009,Eckel10}.
In particular, it requires an account of the competition between the Kondo effect
and the current-induced decoherence~\cite{Paaske_Phys.Rev.B70/2004}
in the (generalized) quantum master equation for the nonequilibrium density operator~\cite{Schoeller_Phys.Rev.B80/2009,Saptsov_Phys.Rev.B86/2012}.

From the present point of view of spectroscopy,
the Kondo effect can be considered as a limit of an inelastic \cot feature at $V=\Delta$ as $\Delta \to 0$,
see \Fig{fig:cot}(a).
Its position is simply $V=0$ at gate voltages sufficiently far between adjacent \set resonances by criterion \eq{eq:naive}.
In particular, for transport spectroscopy of atomic and molecular spin systems the Kondo effect  and its splitting into \cot features~\cite{Heersche_Phys.Rev.Lett.96/2006b,Scott_ACSNano4/2010,Komeda_NatureCommun.2/2011,Frisenda_NanoLett.15/2015}
is very important especially
in combination with strong magnetic anisotropy~\cite{Romeike_Phys.Rev.Lett.96/2006,Leuenberger_Phys.Rev.Lett.97/2006,Romeike_Phys.Rev.Lett.106/2011,Wegewijs11err,Zitko_Phys.Rev.B78/2008,Zitko_NewJ.Phys.11/2009,FernandezRossier09,Zitko_NewJ.Phys.12/2010,Elste_Phys.Rev.B81/2010,Misiorny_Phys.Rev.Lett.106/2011,Misiorny_Phys.Rev.B84/2011,Delgado11,Misiorny_Phys.Rev.B90/2014}.
We refer to reviews on STM~\cite{Ternes_J.Phys.:Condens.Matter21/2009,Brune_Surf.Sci.603/2009,Gauyacq_Prog.Surf.Sci.87/2012,Lounis14} and QD~\cite{Florens_J.Phys.:Condens.Matter23/2011} studies.

\paragraph{``Golden Rule'' $\mathcal{T}$-matrix rates.---\label{sec:tmatrix}}

A second way of arriving at the master equation \eq{eq:cot}
 and the rates in $\vec{W}^{1,1}$
is the so-called $\mathcal{T}$-matrix approach.
In essence, here \cot is regarded as a scattering process:
in the ``Golden Rule'' the next-to-leading order $\mathcal{T}$-matrix~\cite{Bruus_book},
\begin{align}
  \mathcal{T}(E)\approx \H^\T\frac{1}{E - \H-\H^\res}\H^\T + \ldots
  ,
  \label{eq:t-matrix}
\end{align}
is used instead of the coupling $\H^T$,
where $E$ is the scattering energy.
The main shortcoming of this approach is that the $\mathcal{T}$-matrix rates so obtained are infinite.
The precise origin of the divergences was identified in \Ref{Koller_Phys.Rev.B82/2010}
to the neglect of contributions that formally appear in \emph{first-order in $\Gamma$}
but which effectively contribute only in second order
to the stationary state~\cite{Leijnse_Phys.Rev.B78/2008}.
\footnote{These come from so-called secular contributions, involving \emph{off-diagonal} elements of the density matrix in the energy basis, in addition to the diagonal elements, the probabilities.}
By taking these contributions consistently into account~\cite{Koller_Phys.Rev.B82/2010},
finite effective rates\cite{foot:rates} for the master equation for the probabilities are obtained.
In both the \asw and \mbox{$\mathcal{T}$-matrix} approach these contributions are ignored and, instead, 
finite expressions for the rates are obtained only after  \textit{ad-hoc} infinite subtractions~\cite{TurekMatveev02,Koch_Phys.Rev.B74/2006}.
This regularization ``by hand'' can ---and in practice does--- lead to rates \emph{different} from the consistently-computed finite rates,
see \Ref{Koller_Phys.Rev.B82/2010,Begemann10} for explicit comparisons.
These problems have also been related~\cite{Timm_Phys.Rev.B77/2008} to the fact that calculation of stationary transport using a density matrix (occupations) is \emph{not} a scattering problem  ---although it can be connected to it~\cite{Timm11}---
in the following sense: the coupling to the electrodes is never adiabatically turned off at large times (i.e., there is no \emph{free} ``outgoing state'').

As we discuss next, such a consistent first \emph{plus} second order approach is not only technically crucial
but this also leads to additional physical effects  that we measure in \Sec{sec:experiment}.

\subsection{\Qd -- \stm crossover\label{sec:crossover}}

\begin{figure}[t]
	\includegraphics[width=0.99\columnwidth]{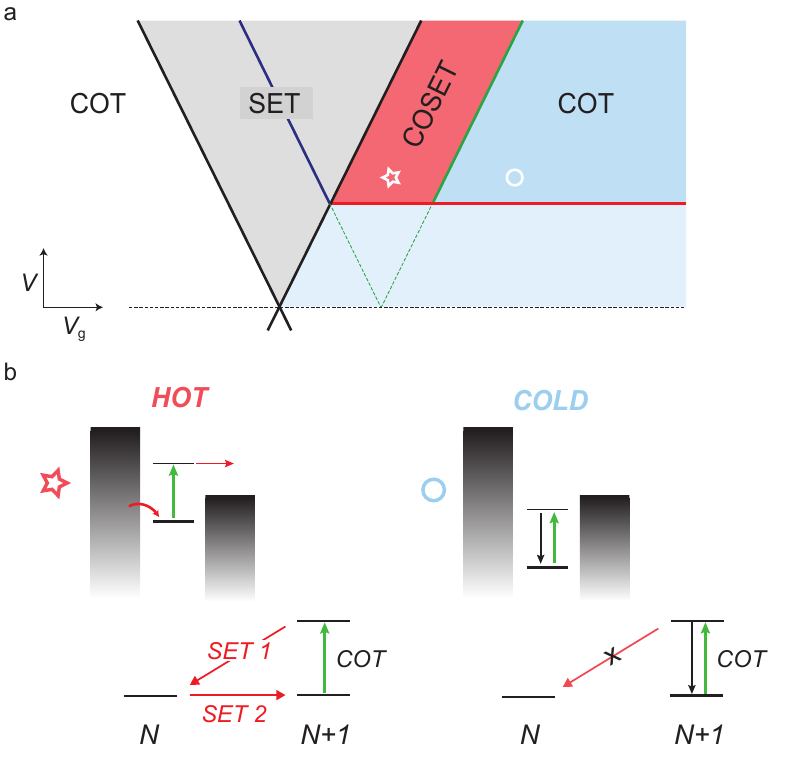}
	\caption{\textbf{Crossover regime between \qd and \stm transport.}
		(a) Same as \Fig{fig:set}(a), now indicating the regimes where the excited state $(N+1,e)$
		relaxes by \cot (blue) or by \set (red).
		(b) Relaxation mechanisms after excitation by \cot:
		\emph{Left panel}:
		relaxation in two steps via ``real'' occupation of charge state $N$.
		\emph{Right panel}:
		when the  process ``SET 1'' is energetically not allowed
		excitation and relaxation proceeds by \cot
                using charge state $N$ only ``virtually''.
	}
	\label{fig:crossover}
\end{figure}

Having reviewed the two prominent, complementary pictures of transport due to ``real'' and ``virtual'' charging,
we now turn to the crossover regime where these two pictures coexist.
This has received relatively little attention, but our experiment in \Sec{sec:experiment} highlights its importance.
As we have seen, despite the fact that charging is only ``virtual'', an energy exchange between molecule and scattering electrons can occur.  
Depending on the energy-level positions, this ``virtual'' tunneling can ``heat'' the molecule so as
to switch on ``real'' charging processes even 
well \emph{outside} the \qd regime.
However, in contrast to real heating,
which leads to smearing of transport features,
this nonequilibrium effect actually results in sharp features in the transport
as a function of bias voltage.
It thus becomes  a new tool for \emph{spectroscopy}.

\subsubsection{\set mirages of \cot excitations\label{sec:coset-exp}}

We first consider the simple case of a single excited state at energy $E_e^{N+1}=E_g^{N+1}+\Delta$ for $N+1$ electrons.
In \Fig{fig:crossover}(a)
we see that the resulting \cot resonance at $V=\Delta$ (red)
connects to the excited-state \set resonance
$\mu_{\L} = \Delta + E_g^{N+1}-E_g^{N}$ (blue), see also \Fig{fig:cot}(a).
The other \set resonance condition for the excited state,
\begin{align}
  \mu_{\R} = \Delta + E_g^{N+1}-E_g^{N}
  \label{eq:coset-res}
\end{align}
defines the green line dividing the inelastic \cot regime $V\geq \Delta$ into two regions shaded red and blue.
In the one shaded blue,
at the point marked with a circle,
the excited state created by a \cot process is stable,
that is,
it cannot decay by a single-electron process since
$ \Delta + E_g^{N+1}-E_g^{N} < \mu_{\R}$.
As shown in the right panel of \Fig{fig:crossover}(b),
the relaxation of this stable state can then only proceed by another \cot process
---via ``virtual'' charging--- and it is thus slow ($\propto \Gamma^2$).
Essentially, this means that the molecule is not ``hot'' enough to lift the Coulomb blockade
of the \emph{excited} state.

In contrast, in the red shaded area,
 at the point marked with a star,
this stability is lost as
$ \Delta + E_g^{N+1}-E_g^{N} > \mu_{\R}$.
Now the relaxation proceeds much faster through a single-electron process (order $\Gamma$)
as sketched in the left panel of \Fig{fig:crossover}(b).
The molecule gets charged for ``real'' (either $N$ or $N+2$) and
quickly absorbs / emits an electron
returning to the stable $N+1$ electron \emph{ground} state,
where the system idles waiting for the next \cot excitation. 
Notably,
this quenching of the excited state takes place far away from the resonant transport regime
in terms of the resonance width,
i.e., violating the linear-response criterion \eq{eq:naive} for being ``off-resonance''.

The enhanced relaxation induced by first-order tunneling,
occurring when moving from the circle to the star in \Fig{fig:crossover}(a),
leads to a change in current
if no other processes (e.g., phonons, hyperfine coupling, etc.) dominate this relaxation channel ($\propto \Gamma$).
As a results, the presence of such a resonance 
signals a ``good'' molecular device, i.e.,
one in which the intrinsic relaxation is small compared to the ``transport-coupling'' $\Gamma$.
We refer to this resonance, first pointed out in \Refs{DeFranceschi_Phys.Rev.Lett.86/2001,Wegewijs01,Golovach_Phys.Rev.B69/2004} and studied further~\cite{Schleser_Phys.Rev.Lett.94/2005,Aghassi_Appl.Phys.Lett.92/2008,Becker_Phys.Rev.B77/2008,Weymann_Phys.Rev.B78/2008,Leijnse_Phys.Rev.B78/2008}, as cotunneling-assisted \set or \coset.

\begin{figure}[t]
\includegraphics[width=0.99\columnwidth]{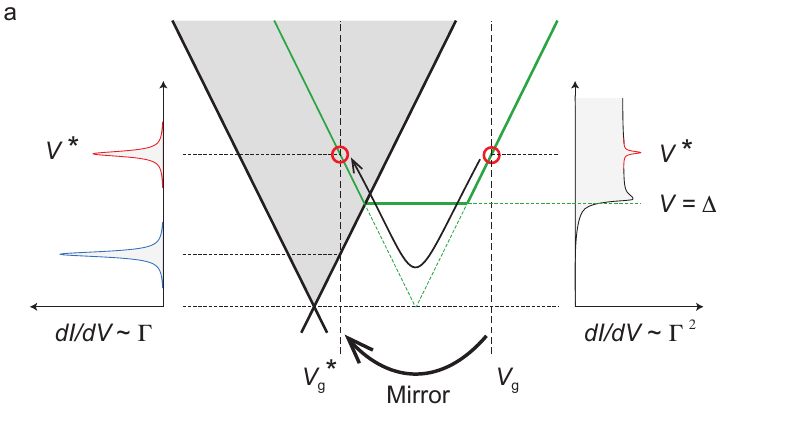}
 \caption{\textbf{Identifying a resonance as a \coset mirage.}
Same situation as \Fig{fig:crossover}(a).
The vertical \didv cut on the right
shows a \cot resonance at $V=\Delta$ and its \emph{mirage} at some bias $V^{*} > \Delta$.
To identify the latter as such, a corresponding \set resonance  must be present
at the mirrored gate voltage $V_{\g}^{*}$,
as in the vertical cut shown on the left.
Note that the indicated construction works for nonsymmetric capacitive coupling.
For symmetric coupling, one can literally mirror the gate-voltage position relative to $\Delta$ on the horizontal axis.
}
 \label{fig:mirage}
\end{figure}

The \coset resonance has both \cot and \set character.
On the one hand, the geometric construction in \Fig{fig:crossover}(a) and \Fig{fig:mirage}
shows that it stems from \emph{the same} excitation as the \cot step at $V=\Delta$.
However, its position $V^{*}$ has the same strongly gate-voltage dependence as a \set resonance,
in contrast to the original \cot resonance at $V=\Delta$.
Yet, the \coset peak requires \cot to appear and its amplitude is relatively weak,
whereas the \set peak is strong and does not require \cot.   
For this reason, the \coset peak can be seen as a \emph{``mirage''} of the \cot excitation and a ``mirror image'' of the 
$(N,g) \to (N+1,e)$ \set peak, 
as constructed in \Fig{fig:mirage}(a).
The resulting mirrored energy conditions can easily be checked in an experiment ---cf. \Fig{fig:13}--- and impose constraints on spectroscopic analysis: 
if  \didv shows a resonance as a function of bias \emph{outside} the \set regime,
a resonance at the mirrored position \emph{inside} the \set regime should be present.

\subsubsection{Connecting \stm and \qd analyses\label{sec:connecting}}

Besides the appearance of \coset mirages, the crossover regime provides 
further important pieces of spectroscopic information
by constraining how \set and \cot spectra
continuously connect as the gate voltage is varied.
This is discussed in \Sec{sec:theory-qd}, \sec{sec:stm} and later on in \Sec{sec:qd-exp},
but we summarize the rules here.
First, only ``equilibrium'' \cot transitions 
---the only ones connecting to an \set resonance as we explained in \Fig{fig:crossover}---
can exhibit a \coset mirage.
Second, excited-excited \cot transitions (i.e., for the same charge state $N+1$)
\emph{never} connect to a corresponding \set feature, as we illustrated in panel (i) of \Fig{fig:cot}(b).
Finally, transitions between excited states with different charge
---visible in the \set regime---
\emph{never} connect to a \cot feature as will be illustrated in \Fig{fig:11}.
These are strict consistency requirements when analyzing the transport spectra in the \set-\cot crossover regime.

\subsubsection{When is transport ``\stm'' ?}

\begin{figure}[t]
  \includegraphics[width=0.99\columnwidth]{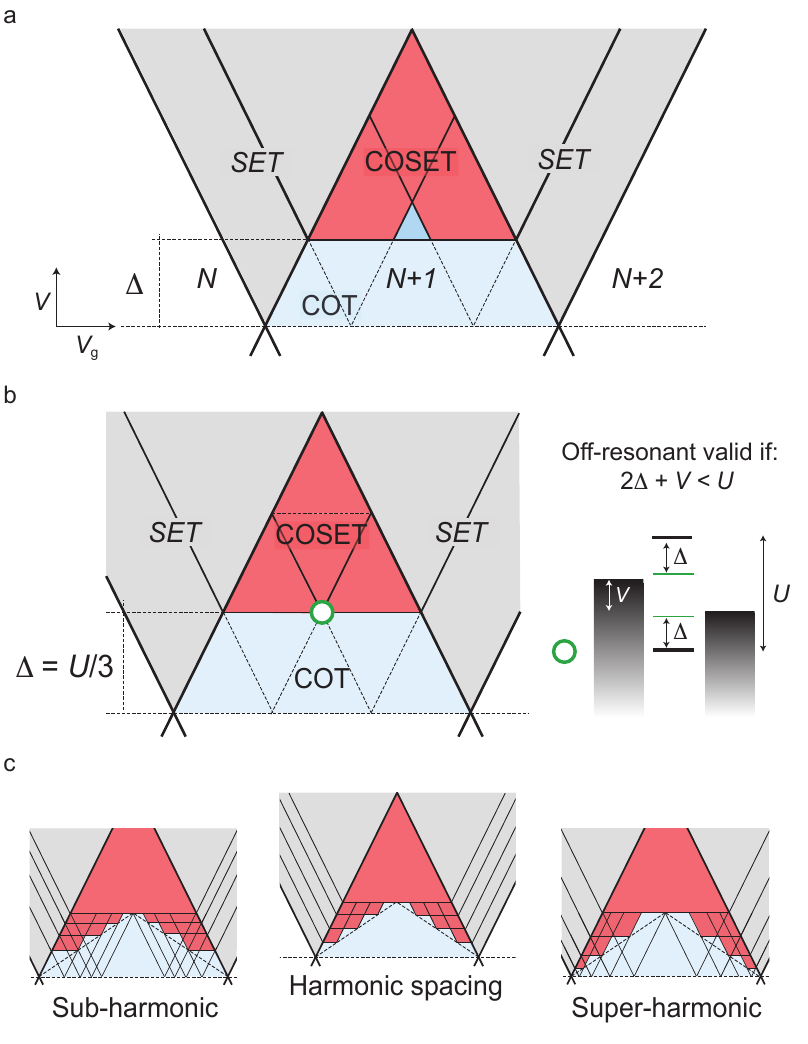}
  \caption{\textbf{Shape and size of the \stm regime.}
    (a)  Same as \Fig{fig:cot}(a) for an excitation $\Delta < U/3$.
    Regions where the \stm approach is valid (fails) are colored blue (red),
    as in \Fig{fig:crossover}.
    In the light blue region $V < \Delta$ there is only elastic \cot
    (dashed black construction lines are not resonances),
    but for $V \gtrsim \Delta$ inelastic \cot does excite the molecule.
    However, the \stm approach only applies in the darker blue triangle
    where both excitation \emph{and relaxation} proceed by ``virtual'' charging.
    This regime is restricted from both sides [\Eq{eq:stableex}]
    and shrinks as $\Delta$ increases.
    (b)
    For $\Delta = U/3$ the \stm picture works only for \emph{elastic} \cot, i.e., for $V < \Delta$,
    which amounts to $5/9 \approx 55 \%$ of the \emph{nominal} \stm regime.
    The inset explains the threshold $\Delta < U/3$:
    for fixed gate voltage at the center ($\medcircle$)
    ---the best-case scenario for the \stm picture to work---
    there is no relaxation by \set as long as the bias satisfies $2 \Delta + V < U$.
    Requiring this to hold at the onset of excitation by \cot, $V=\Delta$, gives the threshold.
    (c)
    Several excitations from a
    {superharmonic} (left),
    {harmonic} (center) and 
    {subharmonic} (right) spectrum for charge $N+1$.
    In the limit of vanishing harmonic energy spacing, the blue region where the \cot picture works approaches $1/3 \approx 33 \%$ of the nominal off-resonant regime.
  }
  \label{fig:validity}
\end{figure}

We are now in the position to determine the region
in which the physical picture of \stm scattering through the molecule of \Sec{sec:stm} applies.
This is illustrated in \Fig{fig:validity}.

The key necessary assumption of the \cot approach ---often not stated precisely---
is that all excited states $(N+1,e)$ that are \emph{accessible} from the ground state $(N+1,g)$
must be ``stable'' with respect to \emph{first-order} relaxation processes:
\begin{align}
  W^{N,N+1}_{g,e}=0 
  \quad 
  \text{and} 
  \quad 
  W^{N+2,N+1}_{g,e}
  =
  0
  .
\label{eq:stableex_rates}
\end{align}
This is the case if the \set condition \eq{eq:set-res} additionally holds
for the \emph{excited} states, i.e., for $i=e$ in \Eq{eq:set-res}:
\begin{align}
  \mu_r < E^{N+1}_e - E^{N}_g 
  \quad 
  \text{and}  
  \quad
  \mu_r > E^{N+2}_g - E^{N+1}_e
  \label{eq:stableex}
\end{align}
for both $r=\L,\R$. We note that in theoretical considerations, it is easy to lose sight of condition \eq{eq:stableex}
when  ``writing down'' an \asw Hamiltonian model (or only $\mathcal{T}$-matrix rates for \cot)
 [\Sec{sec:theory-stm}]
and assuming the couplings to be fitting \emph{parameters} of the theory.\footnote{(In fact, one has to add to condition \eq{eq:stableex} that ``accessible'' means also accessible via nonequilibrium cascades of \cot transitions (``nonequilibrium \cot''),
but we will not discuss this further complication.)}

In \Fig{fig:crossover}(a) we already shaded in light blue the region bounded by the first condition \eq{eq:stableex} where the \cot picture applies.
In \Fig{fig:validity}(a) we now show that the full restrictions 
imposed by both ``virtual''  charge states $N$ and $N+2$ in \eq{eq:stableex}
strongly restrict the validity regime of the \cot approach for states with ``real'' occupations and charge $N+1$.
In \Fig{fig:validity}(b) and its caption
we explain that for any individual excitation $\Delta> U/3$
the \stm picture \emph{always} breaks down
in the sense that it works only for \emph{elastic} \cot, i.e., for $V < \Delta$.
This  amounts to $55 \%$ of the \emph{nominal} \stm regime.

When accounting for several excited states below the threshold $U/3$,
a sizeable fraction of this region must be further excluded.
In \Fig{fig:validity}(c) we construct the regime of validity (blue)
for some example situations.
The shape and size of this validity regime (light blue) depends on the details of the excitation spectrum.
The center panel illustrates that for a harmonic spectrum
the \cot picture in fact applies in only $\sim 33 \%$ of the nominal \stm regime
(i.e., obtained by taking the complement of the \qd regime).
The left and right panel in \Fig{fig:validity}(c)
show how this changes for anharmonic spectra characteristic of quantum spins
with positive and negative magnetic anisotropy, respectively.

In summary,
\qd processes \emph{always} dominate the relaxation of excitations at energy $\Delta > U/3$ populated by \stm excitation because they are ``too hot'':
for such excitations there is no ``deep'' or ``far off-resonant'' regime where considerations based on the \cot picture alone are valid.
For lower-energy excitations, 
$\Delta \leq U/3$,
there is a triangular-shaped region in which one is still truly ``far off resonance'' and excitations are not quenched. The size of that region varies according to \eq{eq:stableex}
and is much smaller than naively expected by extending the linear-response criterion~\eq{eq:naive}.
Although theoretical~\cite{Golovach_Phys.Rev.B69/2004,Aghassi_Appl.Phys.Lett.92/2008,Becker_Phys.Rev.B77/2008,Weymann_Phys.Rev.B78/2008,Leijnse_Phys.Rev.B78/2008, Lueffe_Phys.Rev.B77/2008} and experimental~\cite{DeFranceschi_Phys.Rev.Lett.86/2001,Schleser_Phys.Rev.Lett.94/2005} studies on \coset exist,
this point seems to have been often overlooked and is worth emphasizing.
Experimentally, to be sure that the \stm  picture applies to unidentified excitation
one must at least have an estimate of the gap $U$ and of the level position
or, preferably, a map of the dependence of transport on the level position independent of the bias
as in gated experiment discussed in \Sec{sec:experiment} or STM situations allowing for mechanical gating~\cite{Temirov_Nanotech.19/2008,Toher11,Greuling11a,Greuling11b,Greuling13,Wagner_Prog.Surf.Science90/2015}.

\subsubsection{Stationary state and current at the \qd-\stm crossover}

Due to their hybrid character, \coset mirages do not emerge in a picture of
either ``real'' or ``virtual'' charging alone.
In particular, \set processes are omitted when deriving the \cot rates by means of the \asw transformation [\Sec{sec:stm}],
and, for this reason, that picture cannot account for these phenomena. 
Instead, a way to capture these effects is to extend Eqs.~\eq{eq:set-master} and \eq{eq:cot} to a master equation which simultaneously
includes transition rates 
of leading ($\Gamma$) and next-to-leading order ($\Gamma^2$).
This has been done using the $\mathcal{T}$-matrix approach~\cite{TurekMatveev02,Koch_Phys.Rev.B74/2006}, requiring 
the \textit{ad-hoc} regularization by hand mentioned in \Sec{sec:tmatrix}.
A systematic expansion which avoids these problems is, however, well-known~\cite{Koenig_Phys.Rev.B54/1996,Koenig_Phys.Rev.Lett.78/1997,Koenig_Phys.Rev.B58/1998}.
and we refer to \Refs{Leijnse_Phys.Rev.B78/2008,Koller_Phys.Rev.B82/2010,Gergs17a} for calculation of the rates.

Relevant to our experiment in \Sec{sec:experiment} is that with the computed rates in hand, a stationary master equation needs to be solved to obtain the occupation of the states and from these the current.
We stress that \emph{even when far off-resonance}
---where naively speaking ``the charge is fixed'' to, say, $N+1$---
a description of the transport
requires a model which also includes \emph{both the $N$ and $N+2$ charge states},
together with their relative excitations.
This is essential to correctly account for the relaxation mechanisms that visit these states ``for real'' and not ``virtual''.
\footnote{Note that keeping these states is \emph{not} related to obtaining the correct strength of the couplings for scattering in the \asw Hamiltonian.
Even with the correct values for $J$ and $K$ in \Eq{eq:HA},}
the \coset mirages are missed since $\H^\A$ only accounts for scattering processes.
The minimal master equation required for \stm transport thus takes then the form:
\begin{align}
  \frac{\text{d}}{\text{d}t}
  \begin{bmatrix}
    \vec{P}^{0} \\
    \vec{P}^{1} \\
    \vec{P}^{2} 
  \end{bmatrix}
   = 0 =
   \begin{bmatrix}
    \vec{W}^{0,0} & \vec{W}^{0,1} & \vec{W}^{0,2} \\
    \vec{W}^{1,0} & \vec{W}^{1,1} & \vec{W}^{1,2} \\
    \vec{W}^{2,0} & \vec{W}^{2,1} & \vec{W}^{2,2}
  \end{bmatrix}
  \begin{bmatrix}
    \vec{P}^{0} \\
    \vec{P}^{1} \\
    \vec{P}^{2} 
  \end{bmatrix}
  ,
  \label{eq:full}
\end{align}
where as before $N=0$ for simplicity. 
Here the rates for the various processes change whenever one of the energetic conditions \eq{eq:set-res} and \eq{eq:cot-res} is satisfied. 
Examination of the various contributions in the expression of the rate
matrices~\cite{Leijnse_Phys.Rev.B78/2008} reveals that the following effects are included:
\begin{itemize}[leftmargin=*]
\item
  $\vec{W}^{1,0}$ is a matrix of \set rates
  that change when condition \eq{eq:set-res} is met.
  It also includes $\Gamma^2$-corrections that \emph{shift and broaden} the \set resonance.
\item
  $\vec{W}^{1,1}$ is matrix of both \set and \cot rates.
  The latter change when condition \eq{eq:cot-res} is met.
\item
  $\vec{W}^{2,0}$ and $\vec{W}^{0,2}$ are matrices of \emph{pair-tunneling} rates,
    e.g., $W^{2,0}_{f,i}$
  for transitions between states differing by \emph{two} electrons, $(N,i) \to (N+2,f)$.
  These lead to special resonances discussed in \Sec{sec:break}.
\end{itemize}
The solution of the full stationary master equation \eq{eq:full}
requires some care~\cite{Leijnse_Phys.Rev.B78/2008,Koller_Phys.Rev.B82/2010}
due to the fact that it contains both small \cot rates and large \set rates
whose  interplay produces the \coset mirages.
Even though the 
\emph{(first-order) \set rates} are large,
they have a small ---albeit non-negligible--- effect since, in the stationary situation,
the initial states for these transitions may have only small occupations.
These occupations, in turn, depend on the competition between \emph{all} processes / rates in the stationary limit.
This is the principal reason why one cannot avoid solving the master equation \eq{eq:full}
with both first and second order processes included.

To conclude, \Eq{eq:full} captures the delicate interplay of \qd (\set) and \stm (\cot) processes leading to mirages (\coset).
The appearance of such mirages indicates that \emph{intrinsic} relaxation rates are smaller than \set transport rates ($\propto \Gamma$). 
``Nonequilibrium'' \cot is also included in this approach and the appearance of its additional features in our experiment signals a molecular device with even lower intrinsic relaxation rates, i.e.,
smaller than the \cot relaxation rates ($\propto \Gamma^2$). 

\subsection{Breaking the rules of transport spectroscopy\label{sec:break}}

\begin{figure}[t]
	\includegraphics[width=0.99\columnwidth]{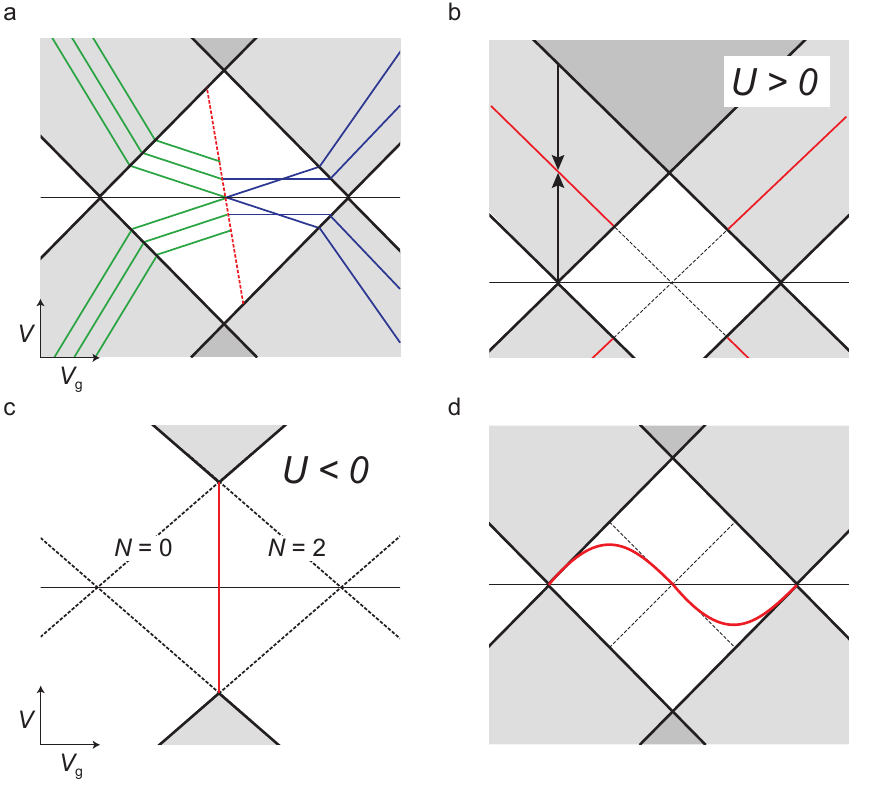}
	\caption{\textbf{Breaking the rules of transport spectroscopy.}
		(a) Effects of nonuniform gate and bias dependence sketched after Fig.~3 of \Ref{Osorio_NanoLett.10/2010}.
		Split singlet-to-triplet inelastic \cot excitations (green) are tuned to degeneracy by a strong $V_{\g}$-dependence.
		Upon crossing the red line, the ground state changes from singlet to triplet, see main text.
		(b-c) Electron-pair tunneling resonance (red) for
		(b) repulsive electron interaction $U>0$
		and
		(c) effectively attractive interaction $U<0$.
		(d) Transport feature (red) due to coherent spin-dynamics
		on a single, interacting orbital coupled to nearly antiparallel ferromagnets.
		Although it looks like a resonance with anomalous gate and bias dependence
		it does not correspond to any state on the system.
		It is instead a sharp amplitude modulation caused by the \emph{orientation} of the accumulated nonequilibrium spin
		relative to the electrode polarization vectors.
	}
	\label{fig:break}
\end{figure}

The above account of the basic rules of transport spectroscopy, although extensive,
is by no means exhaustive.
The key conditions are
\Eq{eq:set-res} and \eq{eq:cot-res},
determining the \emph{resonance positions} as a function of applied voltages.
Readers interested mostly in the application of these rules to a high-resolution transport experiment can skip the remainder of this section and proceed directly
to \Sec{sec:experiment}.
Here, we give an overview of a variety of additional effects that bend or break these rules,
found in experimental and theoretical studies.
In \Fig{fig:break} we sketch a number of transport spectra that cannot be understood from what we have learned in the previous discussion.

\paragraph{Nonuniform level shifts due to voltages.---}
The assumption made so far [\Sec{sec:qd-gate}] that all energy levels are uniformly shifted by applied gate and bias voltages may not be valid in case of local electric field gradients.
In fact, this was already seen in the first experiment on gated \cot spectroscopy of a single-triplet semiconductor dot~\cite{DeFranceschi_Phys.Rev.Lett.86/2001}
due to the change of the confining potential with gate voltage.
In molecular junctions this has also been observed.
Figure~\ref{fig:break}(a) schematizes how the transport spectrum in \Refs{Osorio_NanoLett.10/2010}
displays such effects.
In this case, the \cot resonances can still be identified as weakly gate-dependent resonances,
which is not a trivial issue as the experiments in \Ref{Eliasen_Phys.Rev.B81/2010} show.
However, a qualitatively new and strongly gate-dependent resonance\cite{Osorio_NanoLett.10/2010,Stevanato12} (red line) appears upon ground state change.
Piecing together all the evidence, it was shown
that this effect originates from a change in amplitude of the \cot background,
without requiring the introduction of any additional states into the model.
These effects are included in \Eq{eq:full},
which was shown~\cite{Stevanato12} to reproduce the experimental data of \Ref{Osorio_NanoLett.10/2010} in detail.

\paragraph{Pair tunneling.---}

In all the schematics so far we left out resonances that are caused by electron \emph{pair tunneling}.
These are described~\cite{Leijnse_Phys.Rev.B78/2008,Leijnse_Phys.Rev.Lett.103/2009}
by the rates $\vec{W}^{2,0}$ and $\vec{W}^{0,2}$ included in the master equation \eq{eq:full}.
In \Fig{fig:break}(b) we sketch where these pair-tunneling resonances (red lines) are expected to appear:
their positions are obtained by taking the \emph{bias-averaged positions} of the two \emph{subsequent \set resonances.}
This condition follows by requiring the maximal energy of an electron pair
 in the electrode $r$ to match a corresponding molecular energy change.
For example, for a single orbital at energy $\epsilon$ one obtains
$ 2 \mu_r = E^{N+2}_g -E^{N}_g=  2\epsilon + u$
where $u$ is the charging energy.
This gives a bias window in which pair tunneling 
$N\leftrightarrow N+2$
can contribute to transport,
\begin{align}
  \mu_{\L}  \geq \epsilon + \tfrac{1}{2} u  \geq \mu_{\R} ,
  .
\end{align}
provided that the 
$N$ and / or the $N+2$
state is  occupied.
The effective \emph{charging energy} for each electron is \emph{halved}
since the energy $u$ is available for both electrons together in a single process.
Although small (comparable with \cot) 
its distinct resonance position and shape clearly distinguish the pair-tunneling current  from \set current~\Ref{Leijnse_Phys.Rev.Lett.103/2009}
that dominates in the \qd regime where it occurs.

\paragraph{Electron attraction.---}

Clearly, pair tunneling effects are expected to become important if the effective interaction energy $u$ is attractive~\cite{Haldane_Phys.Rev.B15/1977,Koch_Phys.Rev.Lett.96/2006,Hwang_Phys.Rev.B76/2007,Koch_Phys.Rev.Lett.96/2006,Koch_Phys.Rev.B75/2007,Sela_Phys.Rev.Lett.100/2008}.
Such attraction in fact appears in various systems. In molecular systems this is known as electrochemical ``potential-inversion''~\cite{Evans99}.
In artificial QDs a negative $u$ have been observed experimentally~\cite{Cheng15,Hamo16} in transport spectra of the type sketched in \Fig{fig:break}(c), see also \Ref{Nilsson16}.
Interestingly, in this case the ground state has either $N$ or $N+2$ electrons
and never $N+1$ since starting from $(N,g)$ the single-electron transition energies $E^{N+1}_g-E^{N}_g$ and $E^{N+2}_g-E^{N+1}_g$
 are higher than electron-pair transition energy \emph{per electron}
$(E^{N+2}-E^{N})/2$.
This is also included in the approach \eq{eq:full}, see \Ref{Gergs17a}.

\paragraph{``Coherence'' effects.---}
Finally, we turn to the assumption used in \Sec{sec:theory-qd} that
the molecular state is described by ``classical'' occupation probabilities of the quantum states
(statistical mixture).
For instance, each degenerate spin multiplet is treated as an ``incoherent'' mixture of different spin projections (no quantum superpositions of spin-states states).
Equivalently, the spin has no average polarization in the direction transverse to the quantization axis.

However, when in contact with, e.g., spin-polarized electrodes, such polarization does arise already in order $\Gamma$.
In that case one must generalize \Eq{eq:full} to include off-diagonal density-matrix in the energy eigenbasis.\footnote{(The off-diagonal elements also came into play when going to order $\Gamma^2$, see discussion in \Sec{sec:tmatrix}.)}
In physical terms, this means that one must account for the coupled dynamics of charge, spin-vector and higher-rank spin tensors~\cite{Sothmann10,Misiorny_NaturePhys.9/2013}.  
In the \set regime, such effects can lead to a nearly 100\% modulation of the transport current~\cite{Braun_Phys.Rev.B70/2004,Sothmann_Phys.Rev.B82/2010} due to quantum interference.
This emphasises that~\cite{Koenig_Phys.Rev.Lett.86/2001} \set ---the first order approximation in $\Gamma$--- is not ``incoherent'' or ``classical''
as some of the nomenclature in \Tab{tab:compare} seems to imply.

Similar coherence effects can arise from
orbital polarization in QDs~\cite{Koenig01,Wunsch05,Donarini06,Haertle14,Wenderoth16} and STM configurations~\cite{Donarini12b},
from an interplay between spin and orbital coherence~\cite{Begemann_Phys.Rev.B77/2008,Donarini09,Donarini10},
or from charge superpositions of electron pairs.
Finally,
for high-spin systems coherence effects of tensorial character can arise.
This leads to the striking effect that in contact with ferromagnets (vector polarization)
they can produce a magnetic anisotropy (tensor)~\cite{Baumgaertel11,Misiorny_NaturePhys.9/2013},
see also related work~\cite{Misiorny_Phys.Rev.B86/2012,Oberg_NatureNanotechnol.9/2014,Delgado_Surf.Sci.630/2014,Jacobson_NatureCommun.6/2015}.
An extension of the approach \eq{eq:full} also describes these effects~\cite{Misiorny_NaturePhys.9/2013,Hell15a}.

The perhaps most striking effect of spin-coherence is depicted in \Fig{fig:break}(d):
\set resonances can \emph{split} for no apparent reason~\cite{Baumgaertel11}
 and wander off deep into the \cot regime~\cite{Hell15a} (red line).
Depending on the junction asymmetry, this feature of coherent nonequilibrium spin dynamics can appear as a pronounced gate-voltage dependent current peak
or as a feature close to the linear response regime, mimicking a Kondo resonance, see also~\Ref{Wenderoth16}.
\section{
Spectroscopy of a high-spin molecule\label{sec:experiment}}

In the second part of this paper
we present feature-rich experimental transport spectra as a function of gate-voltage and magnetic field.
Their analysis requires all the spectroscopic rules that we outlined in the first part of the paper.
We show how the underlying Hamiltonian model can be reconstructed from the transport data, 
revealing an interesting high-spin quantum system with low intrinsic relaxation.

The molecule used to form the junction is a $\text{Fe}_4$
single-molecule magnet (SMM)
with formula [Fe$_4$(L)2(dpm)6]~$\cdot$~Et2O where Hdpm is 2,2,6,6-
tetramethyl-heptan-3,5-dione.
Here, H3L is the tripodal ligand
2-hydroxymethyl-2-phenylpropane-1,3-diol, carrying a
phenyl ring \cite{Accorsi_J.Am.Chem.Soc.128/2006}.
After molecular quantum-dot formation,
the device showed interesting isotropic high-spin behavior
and the clearest signatures of \coset to date~\cite{DeFranceschi_Phys.Rev.Lett.86/2001,Schleser_Phys.Rev.Lett.94/2005}
for any quantum-dot structure.\footnote{The device showed no significant anisotropy splittings of spin multiplets in transport, see discussion below.}
Before turning to the measurements and their analysis,
we first discuss specific challenges one faces probing spin-systems
using either \cot or \set spectroscopy.

\subsection{Principles of spin-spectroscopy.
\label{sec:spectro}}

Isotropic, \emph{high-spin} molecules have molecular states labeled by the spin length $S$ and spin-projection $M$.
To detect them two types of selection rules are frequently used in STM and QD studies.
Using these we construct the possible spectroscopic \cot and \set fingerprints that we can expect to measure.

\subsubsection{Spin selection rules for \cot}

Spectroscopy using  \cot conductance as a function of magnetic field $B$ (``spin-flip spectroscopy''\cite{Heinrich_Science306/2004})
has been a key tool in both STM and break-junction studies.
This approach assumes that ``virtual'' charging
processes dominate. These processes involve two electrons for which the selection rules \eq{eq:rules-cot} apply.

However, for high-spin molecules considered here, 
there can be multiple spin-spectrum assignments that fit the same \cot transport spectrum.
An indication for this is that in the present experiment some of the spectra are very similar
to those of entirely different nanostructures~\cite{Gaudenzi_NanoLett.16/2016}.

\begin{figure}[t!!!]
\includegraphics[width=0.99\columnwidth]{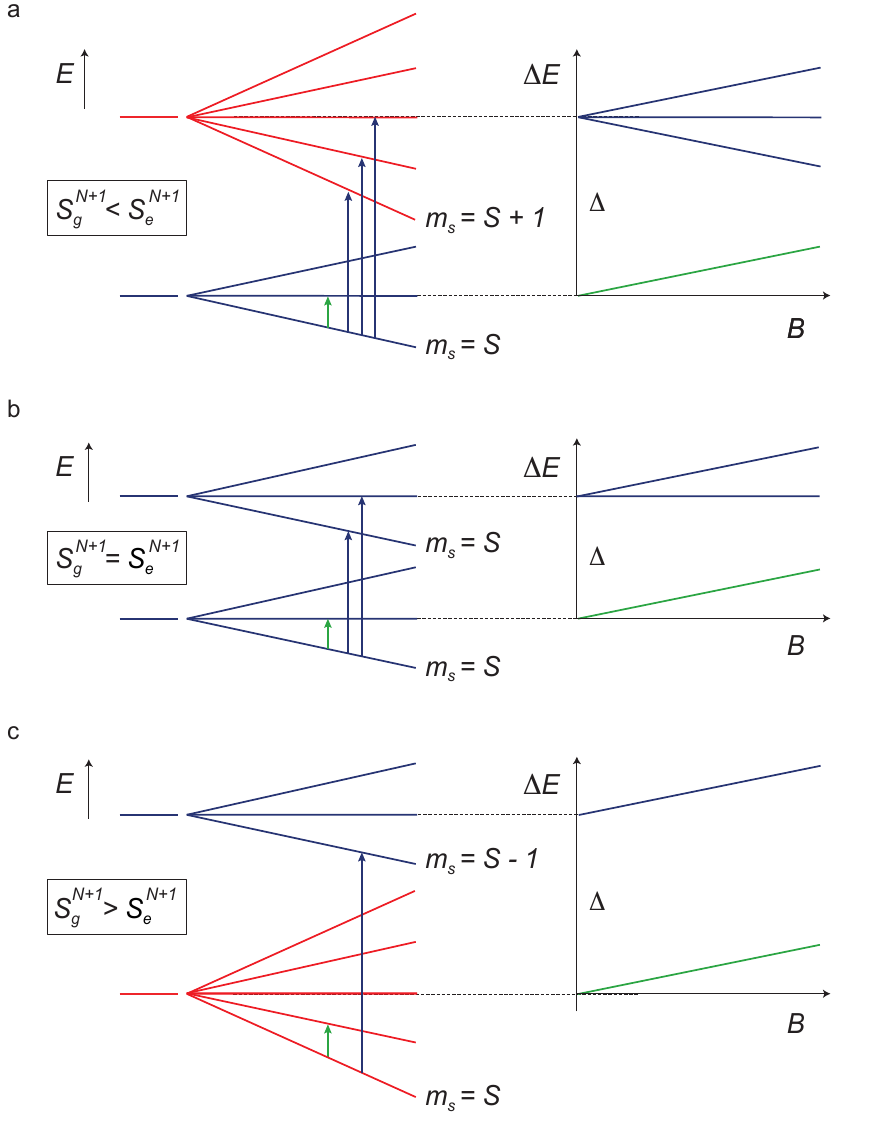}
 \caption{\textbf{\Cot spectroscopy of a high-spin molecule.}
The left panels in (a)-(c) show \cot transitions between energy levels ($E$) vs. a magnetic field $B$.
The right panels show the corresponding transport spectra, i.e., the resonant bias positions in \didv matching an energy difference ($\Delta E$).
(a) If the spin increases upon excitation, $S_e^{N+1} = S_g^{N+1}+1$,
there is a \emph{three-fold} splitting of the transport-spectrum (blue) starting at $V=\Delta$ for $B=0$
due to the transitions to the excited
multiplet.
The ground
multiplet gives a line (green) starting at $V=0$ and increasing with $B$
if $S_g^{N+1} \geq 1/2$.
Only for $S_g^{N+1}=0$ this green line is \emph{missing}.
(b)
If the spin length does not change upon excitation, $S_e^{N+1} = S_g^{N+1}$,
the excited multiplet appears in the transport spectrum
through a \emph{double} line starting at $V=\Delta$.
The ground
multiplet gives a line (green) starting at $V=0$ and increasing with $B$
if $S_g^{N+1} \geq 1/2$.
Clearly, for $S_g^{N+1} = 0 = S_e^{N+1}$ the $B$-dependent lines are \emph{missing}.
(c)
If the spin length decreases upon excitation, $S_e^{N+1} = S_g^{N+1}-1$,
the excited multiplet appears in the transport spectrum
through a single line (blue) starting at $V=\Delta$, increasing with $B$.
Since in this case the ground spin $S_g^{N+1}$ is always nonzero,
there is an intra-multiplet line (green) starting at $V=0$.
}
 \label{fig:spectro}
\end{figure}

To see how this comes about
we construct in \Fig{fig:spectro}(a)-(c) the three possible different fingerprints
that two spin-multiplets can leave in the \cot transport spectrum based on selection rules \eq{eq:rules-cot} alone.
For simplicity, we assume that all processes
start from the ground state $(N+1, g)$, i.e., in the ``equilibrium'' situation discussed in \Sec{sec:cot}.
This figure shows that one can determine only whether the spin value changes  by 1
or remains the same upon excitation, but not on the \emph{absolute} values of the spin lengths (unless the ground state has spin zero).

\subsubsection{Spin selection rules for \set -- spin blockade}
A second key tool in the study of spin effects is
the transport in the \set regime~\cite{Hanson_Rev.Mod.Phys79/2007,Heersche_Phys.Rev.Lett.96/2006a,Grose_NatureMater.7/2008,Burzuri_Phys.Rev.Lett.109/2012,Misiorny_Phys.Rev.B91/2015,Burzuri_J.Phys.:Condens.Matter27/2015}.
This provides additional constraints that reduce the nonuniqueness in the \cot spin-assignment.

\begin{figure}[t]
  \includegraphics[width=0.99\columnwidth]{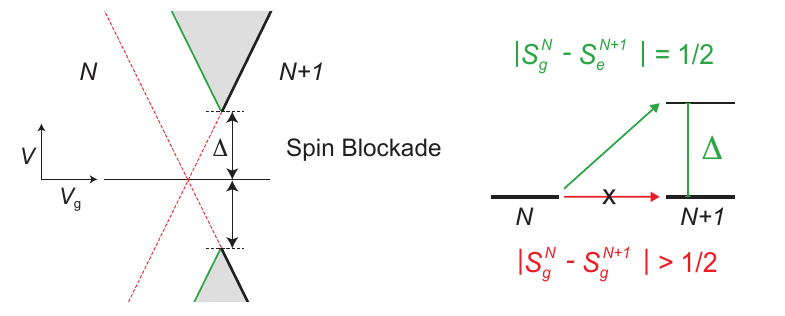}
  \caption{\textbf{\Set spectroscopy of a high-spin molecule: spin blockade.}
    If ground state transitions are spin-forbidden, $|S^{N+1}_g - S^{N}_g | > 1/2$,
    then \set transport is suppressed (red dashed cross).
    Transport sets in only when a finite bias makes the lowest \emph{spin-compatible} excitation energetically accessible.
    This can be either and $N+1$ state with $|S^{N+1}_e - S^{N}_g | = 1/2$ (shown)
    or an $N$ electron state with $|S^{N+1}_g - S^{N}_e | = 1/2$ (not shown).
  }
  \label{fig:spin_blockade}
\end{figure}

In the \set regime, the linear-transport part is governed by
the transition between the two ground-state multiplets with different charge, $(N,g)$ and $(N+1,g)$,
for which selection rules \eq{eq:rules-set} hold.
As sketched in \Fig{fig:spin_blockade}, if linear \set transport is observed, 
then the ground-state spin values are necessarily linked by
\begin{align}
        | S^{N+1}_g - S^{N}_g | = \tfrac{1}{2}
  .
  \label{eq:nospinblock}
\end{align}
This constraint, used in~\Refs{Zyazin_NanoLett.10/2010,Osorio_NanoLett.7/2007,Burzuri_Phys.Rev.Lett.109/2012},
restricts the set of level assignments inferred through \cot spectroscopy on each of the \emph{two} subsequent charge states,
by fixing the relative ground state spins $S^{N}_g$ and $S^{N+1}_g$. 
Their absolute values remain, however, undetermined, unless one of two happens to be zero.
Arguments based on the presence of the 
additional spin-multiplets can then be used to motivate a definite assignment of spin values.

Molecules for which \Eq{eq:nospinblock} fails
can be identified by a clear experimental signature:
the \set transport is blocked up to a finite bias as explained in \Fig{fig:spin_blockade}.
Such \emph{spin-blockade} has been well-studied experimentally~\cite{Huettel_Europhys.Lett.62/2003,Ciorga_Phys.Rev.B61/2000,Johnson_Phys.Rev.B72/2005}
and theoretically~\cite{Weinmann_Europhys.Lett.26/1994,Weinmann_Phys.Rev.Lett.74/1995,Weinmann_Lect.NotesPhys.630/2003,Romeike_Phys.Rev.Lett.96/2006_TranspSpetr,Romeike_Phys.Rev.B75/2007} and finds application  in spin-qubits (``Pauli-spin blockade'').
It has been reported also for a molecular junction~\cite{Osorio_NanoLett.7/2007}.

Clearly, when several excited spin multiplets/charge states are involved, both the \set and \cot spin-spectroscopy become more complex.
However, selection rules similar to \Eq{eq:nospinblock} also apply to \emph{excited} states and
thus ``lock'' the two spin spectra together.  
In addition, the nonequilibrium occupations of the states
contribute to further restricts\cite{foot:internalsplit} the set of possible spin-values
as we will now illustrate in our experimental spectroscopic analysis.

\subsection{Molecular junction fabrication}

Molecular junctions are produced starting from 
a three-terminal solid-state device~\cite{Oneill_Appl.Phys.Lett.13/2007,Gaudenzi_Appl.Phys.Lett.106/2015}
consisting of an oxide-coated metallic local gate electrode with a thin gold nanowire deposited on top. 
On such a device, a low-concentration solution of molecules  ($\sim 0.1$ mM) is drop-casted.     
The nanowire is then electromigrated at room temperature and allowed to self-break~\cite{Oneill_Appl.Phys.Lett.13/2007}
so that a clean nanogap is formed, with a width of $\approx 1.5$~nm.  
The solution is evaporated and the electromigrated junctions are cooled down
in a dilution fridge ($T_\text{base} \approx 70$~mK) equipped with a vector magnet and low-noise electronics.
All the measurements are performed in a two-probe scheme either by applying a DC bias $V$ and recording 
the current $I$ or by measuring $dI/dV$ with a standard lock-in AC modulation of the bias.

A molecular junction as sketched in \Fig{fig:1}(b)
is formed when a molecule
 physisorbs\cite{foot:physisorption} on the gold leads,
and thus establishes a tunneling-mediated electrical contact.
The presence of the molecule in the junction is signaled by 
large \set transport gaps $U$ exceeding $100$ meV
and low-bias inelastic \cot fingerprints.
Numerous molecular systems have been investigated in this configuration~\cite{Heersche_Phys.Rev.Lett.96/2006a,
	Osorio_NanoLett.10/2010,
	Osorio_NanoLett.7/2007, 
	Frisenda16,
	Burzuri_J.Phys.:Condens.Matter27/2015,
	Burzuri_Phys.Rev.Lett.109/2012,
	Zyazin_NanoLett.10/2010,
	Misiorny_Phys.Rev.B91/2015,
	Haque_J.Appl.Phys.109/2011,
	Henderson_J.Appl.Phys.101/2007,
	Jo_NanoLett.6/2006,	
	Parks_Science328/2010,
	Candini_NanoLett.11/2011,
	Urdampilleta_NatureMater.10/2011,
	Vincent_Nature488/2012,
	Frisenda_NanoLett.15/2015,
	Gaudenzi_NanoLett.16/2016}.

As a side remark, the fact that we do not observe pronounced magnetic anisotropy effects
is not unexpected: the formation of a molecular junction may involve surface interactions.
In several cases previously studied
clear spectroscopic signatures of the ``bare'' molecular structure (before junction formation),
such as the magnetic anisotropy~\cite{Jo_NanoLett.6/2006,Zyazin_NanoLett.10/2010,Parks_Science328/2010},
were observed also in junctions.
However, depending on the mechanical and electrical robustness of the molecule,
this and other spin-related parameters may
undergo quantitative~\cite{Burzuri_J.Phys.:Condens.Matter27/2015,Misiorny_Phys.Rev.B90/2014,Kahle_NanoLett.12/2012,Voss09}
or qualitative changes~\cite{Burgess_NatureComm.6/2015,Gaudenzi_NanoLett.16/2016}
and  sometimes offer interesting opportunities for molecular spin control~\cite{Heinrich_Nat.Phys.12/2013}.
Image-charge stabilization effects, for example, can lead to entirely new spin structure
such as a singlet-triplet pair~\cite{Osorio_NanoLett.7/2007, Perrin_NatureNanotechn.8/2013} on opposite sides of a molecular bridge.

\subsection{Characterization of spin states in adjacent redox states}

\begin{figure*}[t]
 \includegraphics[width=0.99\textwidth]{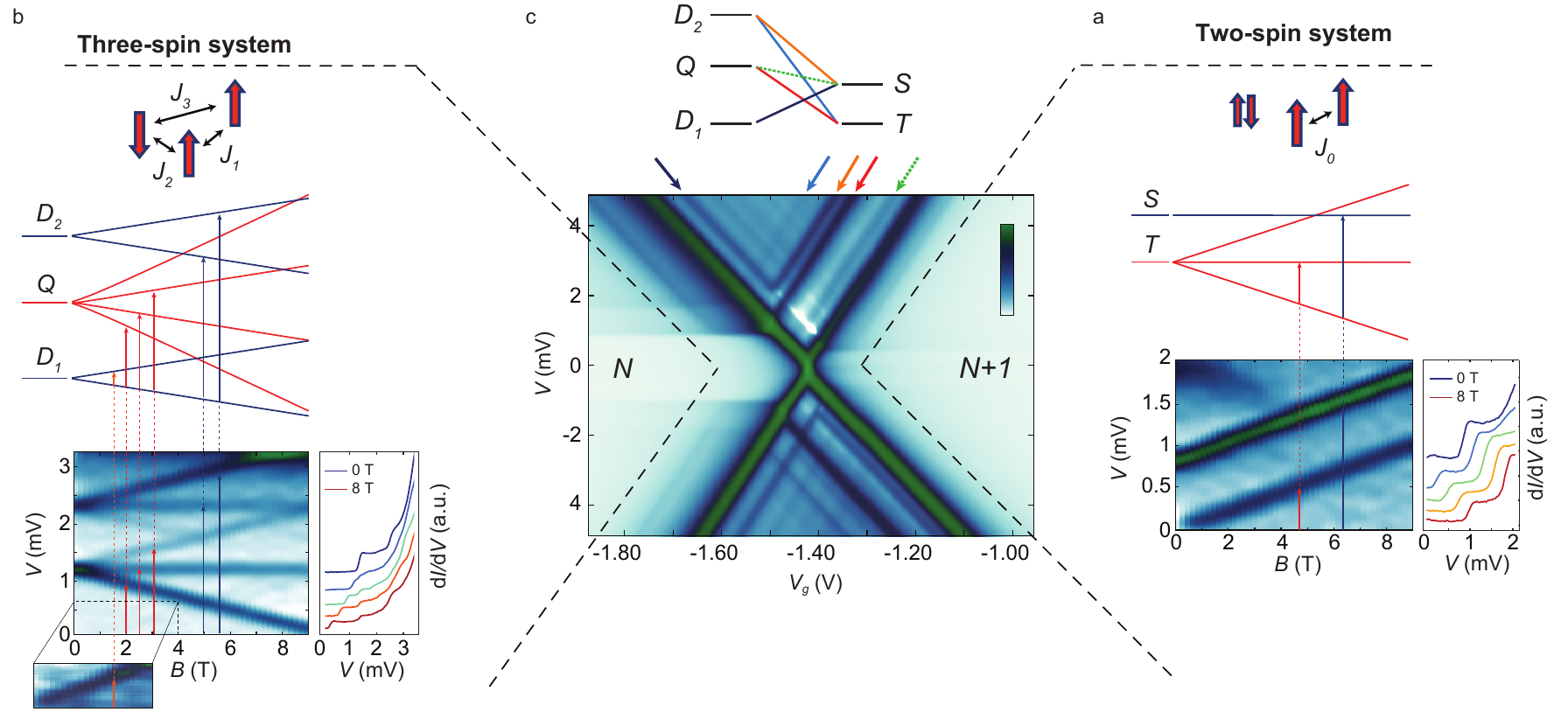}
 \caption{
   \textbf{The characteristics regimes of a complex molecular spin system.} 
   \didv color map (stability diagram) of a high-spin molecular dot.
   The gate electrode allows to electrostatically vary the dot's chemical potential.
   This scanning in energy space grants access to ``real'' charging (mixed-valence, middle)
   as well as the ``virtual'' cotunneling transport regimes (far left and right side, fixed charge).
   Between the two regimes, the hybrid \coset regime is visible where excitation (relaxation) is dominated by \cot (\set).
   \cot spectroscopy at $V_{\g}=-1.81$~V, in (b), ($V_{\g}=-1.25$~V, in (a)) in magnetic field
   reveals the presence of a three (two) spin system with specific ferro-/antiferromagnetic exchange couplings.
   The color-coded arrows indicate the transitions between the different spin multiplets of the three- and two-spin systems.
 }
 \label{fig:10}
\end{figure*}

We now turn to the analysis of the feature-rich transport spectrum anticipated in \Fig{fig:1}(c)
and reproduced in \Fig{fig:10}.
It consists of two \stm regimes on the left and right with fixed charge states
---provisionally labeled $N$ and $N+1$---
and a \qd regime in the center surrounded  by a significant crossover regime.

\subsubsection{\Stm analysis \label{sec:stm-exp}}
We first separately identify the electronic spectrum for each of the two accessible charge states  $N$ and $N+1$
using the \stm approach discussed in \Sec{sec:stm}.

In \Fig{fig:10} (a) we show the \didvv color map and the corresponding \didv steps for fixed $V_{\g}=-1.25$~V as a function of magnetic field, $B$.
Two steps (peaks in \didvv) starting from $V \approx 0$ meV and $V = 0.78$ meV at $B = 0$ T shift upward in energy and parallel to each other as the magnetic field increases. 
In the standard \cot picture each step signal to the opening of an
inelastic transport channel through the molecule. Transport takes place 
via virtual charging involving a real spin-flip excitation with selection rules on spin-length $\Delta S = 0,1$ and magnetization $\Delta M = 0, \pm 1$.
The charge of the molecule remains fixed, and is labeled $N+1$.
The shift in magnetic field of both steps indicates a nonzero spin ground state multiplet with spin $S^{N+1}_{g}$.
According to \Sec{sec:spectro}, the presence of only one other finite-bias excitation 
shifting in magnetic field relates the spin-values as $S^{N+1}_{g} = S^{N+1}_{e} + 1$,
but leaves their absolute values undetermined. 

As we will see later, other spectroscopic information constrains the ground spin to be a triplet $T$,
$S^{N+1}_{g} = 1$,
with a singlet excited state labeled $S$. From the \cot excitation voltage, a 
ferromagnetic (FM) interaction energy $J=0.78$~meV can be extracted. 
Such type of excitation has been seen in other molecular structures~\cite{Osorio_NanoLett.7/2007,Roch_Phys.Rev.Lett.103/2009,Gaudenzi_NanoLett.16/2016,Esat16a}.
Spectra of this kind have also been obtained earlier in other quantum-dot heterostructures,
such as few-electron single and double quantum dots,
albeit typically characterized by smaller and antiferromagnetic couplings~\cite{Sasaki_Nature405/2000, DeFranceschi_Phys.Rev.Lett.86/2001}.

We now change the gate voltage to more negative values so that the molecule is oxidized $N+1 \to N$,
i.e., we extract exactly one electron from the molecule.
This can be inferred from the \set transport regime that we traverse along the way.
In this new charge state we perform an independent \stm spectroscopy.
In \Fig{fig:10}(b) we show the \didvv for $V_{\g} = -1.81$~V as a function of the magnetic field $B$ 
with corresponding \didv line cuts. At $B = 0$ T two sets of peaks in \didvv appear at $V = 1.2$ meV and $V = 2.26 $ meV
and split each in three peaks at higher magnetic fields. A weak excitation shifting upwards in $B$ from $V = 0$ V is also present.  
With the help of \Fig{fig:spectro}(a) the weak excitation and the first set of peaks are associated to
$S^{N}_{e} = S^{N}_{g} + 1$,
while the second set, corresponding instead to the spectrum depicted in \mbox{\Fig{fig:spectro}(b)},
fixes the spin to
$S^{N}_{e'} = S^{N}_{g}$.
The crucial information provided by the clear absence of spin blockade in the intermediate \set regime
eventually constrains $S^{N}_g$ to $1/2$ or $3/2$ according to \eq{eq:nospinblock}. 
The only two spin configurations compatible with the observations are therefore:  
a ground doublet $D_1$, an excited quartet $Q$ and a second doublet $D_2$ or,
alternatively,
a ground quartet, an excited sextuplet and a quartet.
 As we will see in the next section, the latter can be rigorously 
ruled out by analyzing the \set spectrum.  

The presence of the excited quartet state $Q$ implies that the charge state $N$
is a \emph{three-spin} system, 
$N=3$, as sketched in 
the top panel of \Fig{fig:10}(b).
The system with one extra electron in \Fig{fig:10}(a)
is thus
actually a $N+1=4$ electron system with one closed shell, as sketched in the figure.
Upon extraction of an electron,
the spectrum of the molecular device changes drastically, transforming
from a ferromagnetic \emph{high-low} spin spectrum for $N+1=4$
into a nonmonotonic \emph{low-high-low} spin excitation sequence for $N=3$.
The spin-excitation energies extracted from the two independent \cot analyses are:
\begin{equation}\label{eq:energies_3}
\left\{
	\begin{aligned}
	&E_{Q}-E_{D_1}=1.2~\text{meV}
	\\
	&E_{D_2}-E_{D_1}=2.26~\text{meV}
	\end{aligned}
\right.
	\quad
	\text{for}
	\quad
	N=3
	,
\end{equation}
and
\begin{equation}\label{eq:energies_4}
	E_{S}-E_{T}=0.78~\text{meV}
	\quad
	\text{for}
	\quad
	N+1=4	
	.
\end{equation}

These energy differences provide the starting point of a more atomistic modeling 
of the magnetic exchanges in the two charge states.
We stress that for the transport spectroscopy this is not necessary and it goes beyond the present scope.
We only note that
while the $N+1=4$ state requires only one fixed 
ferromagnetic exchange coupling $J_0=E_{S}-E_{T}$ [\Fig{fig:10}(a)]
together with the assumption that two other electrons occupy a closed shell;
the $N=3$ spectrum requires, in the most general case, three distinct exchange couplings 
between the three magnetic centers [\Fig{fig:10}(b)].
These relate to the two available energy differences through $E_{Q} - E_{D_{1}} = (J_{1} + J_{2} + J_{3})/2 + X/2$
and $E_{D_{2}} - E_{D_{1}} = X$ to a complicated function $X(J_{1}, J_{2}, J_{3})$. 
Since this involves three unknowns for two splittings, only microscopic symmetry considerations 
or detailed consideration of the transport current magnitude
are needed to uniquely determine the microscopic spin structure.

This type of microscopic modeling has proven successful in many instances,
see \Ref{Ternes_NewJ.Phys.17/2015} and references therein.
However, the underlying assumptions on localized spins and fixed charge occupations
can only be made when sufficiently far away from resonance,
i.e., such that \coset does not take place
as expressed by conditions \eq{eq:stableex_rates} and \eq{eq:stableex}.

\subsubsection{\Qd analysis\label{sec:qd-exp}}
\begin{figure}[t]
  \includegraphics[width=0.99\columnwidth]{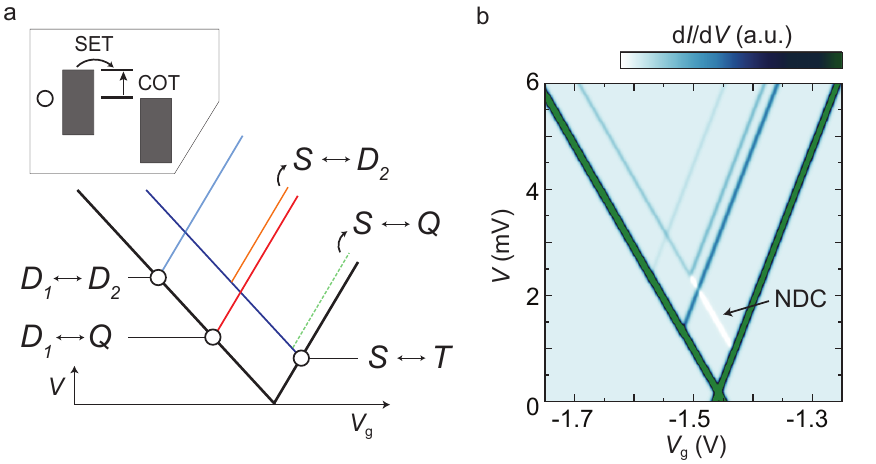}
  \caption{
    \textbf{Connecting the \stm and \qd analyses.}
    (a) 
    Schematic of \cot and \set excitations observed in \Fig{fig:10}.
    \Set transitions between a \emph{ground and excited} state (blue, red) reach the boundary of the \qd regime
    at the black circle from where they continue horizontally as a \cot excitation.
    The inset depicts the chemical potential configuration at such a black circle where \cot and \set connect.
    The \set transitions between \emph{two excited} states (orange, green) do not connect to any \cot excitation.
    (b) 
    \Set transport spectrum computed using the master equations \eq{eq:set-master}-\eq{eq:set-current}.
    The energies are extracted independently from the two \cot spectra in \Fig{fig:10}
    and the capacitive parameters
    $\alpha_{\g}=0.012$,
    $\alpha_{\L}=\alpha_{\g}/0.6$
    $\alpha_{\R}=\alpha_{\g}/0.4$ are fixed by the observed slopes of the \set lines
    [cf. \Fig{fig:set}(a)],
    leaving the tunnel rates \eq{eq:rates_T}-\eq{eq:rates_S} as adjustable parameters.
    The broadening of the \didv peaks in the experiment is due to tunneling,
    $\Gamma \approx 4.6 \text{ K}  \sim 0.4 \text{ meV}$ (FWHM),
    rather than temperature,
    $T            \approx 70 \text{ mK} \sim 6 \text{ $\mu$eV}$.
    %
    %
    \Eq{eq:set-master}-\eq{eq:set-current} do not include this $\Gamma$-broadening
    and we crudely simulate it by an effective higher temperature
    $T^{*} = 270 \text{ mK}           \sim 23 \text{ $\mu$eV}$.
    %
    %
    %
    The master equations \eq{eq:set-master}-\eq{eq:set-current} are valid
    for small effective tunnel coupling  $\Gamma^{*} \ll T^{*}$,
    which only sets the overall scale of plotted \set current
    and not the relative intensities of interest.
    The caption to \Fig{fig:15} explains that
    $\Gamma^{*}$ should not be adjusted to match the larger experimental current magnitude.
  }
  \label{fig:11}
\end{figure}

Using the ability to control the energy levels with the gate,
the \cot analysis can be complemented by a \set spectroscopy in the central part 
of \Fig{fig:10}(c).
Here, 
``real'' charging processes dominate.
For example,
starting from the ground  state $D_1$,
addition of a single electron
leads to occupation of the  $T$ ground state.
This is evidenced by the clear presence of a \set regime of transport down to the linear-response limit.
Inside the \qd regime additional lines parallel to the edges of the cross appear as well.
As we explained in \Fig{fig:set},
these correspond to ``real'' charging processes where excess (deficit) energy is used to excite (relax) the molecule.
These additional lines,
schematized for our experiment in \Fig{fig:11}(a),
fall into two categories according to the criteria:
\begin{enumerate}[label=(\alph*)]
\item Lines terminating at the boundary of the \set regime
  correspond to the \emph{ground} $N$ to excited $N\pm1$ transitions or \emph{vice versa}.
\item Lines that never reach the \set boundary,
  but terminate inside the \set regime at a line parallel to this boundary.
  These correspond to \emph{excited} $N$ to excited $N\pm1$ transitions.
  Their earlier termination indicates that that the initial excited state must become first
  occupied through another process.
  The line \emph{at which} it terminates corresponds to the onset of this ``activating'' process. 
\end{enumerate}

In \Fig{fig:10}(c) and \Fig{fig:11}(a)
the \set transitions
$D_1 \leftrightarrow S$, $Q \leftrightarrow T$ and $D_2 \leftrightarrow T$ fall into category (a),
while the $D_2 \leftrightarrow S$ and $Q \leftrightarrow S$ transitions belong to (b). 
Due to the large difference in spin-length values of the spin-spectra
the latter transition, marked in dashed-green,
is actually forbidden by the selection rules \eq{eq:rules-set}.
Following this line, we find that it terminates at a strong negative differential conductance (NDC) feature
(white in the stability diagram in \Fig{fig:10})
 marking the onset of the transition $D_1 \leftrightarrow S$.

To test our earlier level assignment based \cot, we now compute the expected \set transport spectrum
the first-order ($\Gamma$) master equations \eq{eq:set-master}-\eq{eq:set-current}
and by adjusting the result, we extract quantitative information about the tunnel coupling.
The model Hamiltonian is constructed 
from the energies \eq{eq:energies_3}-\eq{eq:energies_4} and their observed spin-degeneracies.
Assuming that spin is conserved in the tunneling,
the rates between magnetic sublevels are fixed by Clebsch-Gordan spin-coupling coefficients~\cite{Romeike_Phys.Rev.Lett.96/2006,Romeike_Phys.Rev.B75/2007} 
incorporating both the \set and \cot selection rules Eqs.~\eq{eq:rules-set} and \eq{eq:rules-cot}.
The tunnel parameters in units of an overall scale $\Gamma^{*}$ are adjusted
to fit the relative experimental intensities:
\begin{equation}\label{eq:rates_T}
\hspace*{-10pt}
\left\{
	\begin{aligned}
	&\Gamma_{D_1,T} =1.0 \, \Gamma^{*},    
	\\
	&\Gamma_{Q,T} = 1.0 \, \Gamma^{*},
	\\
	&\Gamma_{D_2,T}= 0.25 \, \Gamma^{*}
        \quad \text{(weak intensity)},
	\end{aligned}
\right.
\end{equation}
and
\begin{equation}\label{eq:rates_S}
\left\{
	\begin{aligned}
	&\Gamma_{D_1,S} = 0.5 \, \Gamma^{*},
	\quad\text{(NDC)},
	\\
	&\Gamma_{Q,S} = 0.0  
	\quad\text{(spin-forbidden)},
	\\
	&\Gamma_{D_2,S}=  1.0 \, \Gamma^{*}. 
	\end{aligned}
\right.
\end{equation}
Their relative magnitudes provide further input the further microscopic modeling of the 3-4 spin system
mentioned at the end of \Sec{sec:stm-exp}.
As shown in \Fig{fig:11}(b), 
the \emph{\qd (\set) part} of the experimental conductance in \Fig{fig:10}(c),
as schematized in \Fig{fig:11}(a) is reproduced in detail.
This includes transitions exciting the molecule from its ground states,
but also a transition between excited states.\cite{foot:greenline}
The NDC effect is explained in more detail later on together with the full calculation in \Fig{fig:14}.

\subsubsection{Connecting the \stm and \qd analyses}

As discussed
in \Fig{fig:cot}-\fig{fig:crossover}
 and indicated in \Fig{fig:11}(a)
the \set excitations corresponding to the ground $N$ to excited 
$N\pm1$ transitions connect continuously to the \cot excitations.
Those
corresponding to two excited states, each of a different charge state,
has no corresponding \cot excitation to connect to.
In this sense, the \set spectrum effectively ties the two separately-obtained \cot 
spin spectra and allows a consistency check on their respective level assignments,
cf. \Sec{sec:connecting}.

For instance, from the fact that the $Q \leftrightarrow T$
transition is clearly visible ---marked red in \Fig{fig:10}(c)---
we conclude that the first excited multiplet of the $N$ charge state 
\emph{cannot} be a sextuplet ($S=5/2$)
since such \set transition would be
spin-forbidden and thus weak.
Another example is given by the presence of the
$S \leftrightarrow D_2$ \set transition [orange in \Fig{fig:10}(c)], which implies
that the second excited multiplet of the $N$ charge state \text{cannot} be a {quartet}. 
The fact that this transition does not continue 
into any of the \cot ones is also consistent with its excited-to-excited character.

These two exclusions considerations were anticipated in \Sec{sec:stm-exp}
and are crucial for our \stm assignment 
in the three-electron state
and has now allowed us to reverse-engineer the effective many-electron molecular Hamiltonian.
With this in hand, we turn to the main experimental findings
and investigate the "nonequilibrium" \cot through the molecule [\Sec{sec:pump-probe}]
and the crossover regime where ``real'' and ``virtual'' tunneling nontrivially compete
in the relaxation of spin excitations [\Sec{sec:coset}].

\subsection{Pump-probe spin spectroscopy by nonequilibrium electron \cot \label{sec:pump-probe}}
\begin{figure}[t]
  \includegraphics[width=0.99\columnwidth]{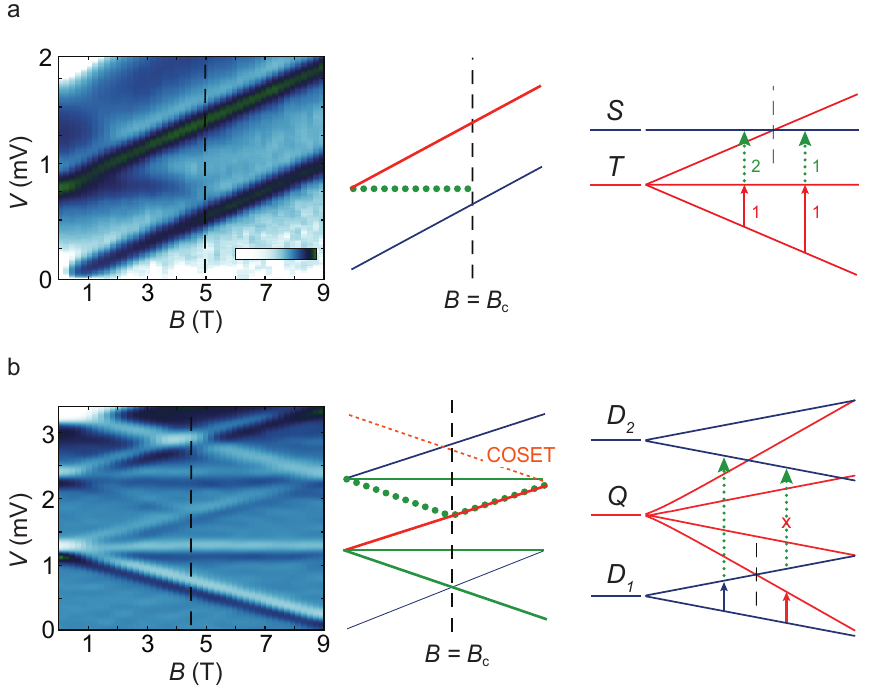}
  \caption{
    \textbf{Nonequilibrium spin pumping and locking mechanism.}
    In contrast to \Fig{fig:10}, we highlight here the transitions that are involved in the spin pumping process (green dotted lines).
    (a) 
    \didvv spectra measured as a function of $B$-field
    at $V_{\g} = - 1.32$~V in the $N+1=4$ charge state. 
    The $T \leftrightarrow S$
    nonequilibrium spin-excitation shows up as a weak, field-independent step vanishing at higher field. 
    For $B < B_c$  the intra-triplet transition (red arrow) requires lower energy than the "nonequilibrium" 
    $T \leftrightarrow S$ transition. 
    For $B > B_c$, the intra-triplet is unlocked (activated) at an energy higher than the 
    $T \leftrightarrow S$ and only one transition of the cascade is visible. 
    (b)
    \didvv spectra measured as a function of $B$-field
    at $V_{\g} = - 1.76$~V in the $N=3$ charge state. 
    Here the nonequilibrium excitation has a negative slope. For $B < B_c$ the excited state of the 
    ground-state doublet $D_1$
    is populated enough to promote a second, nonequilibrium excitation to the excited doublet $D_2$
    (green dotted line). As $B > B_c$ the $D_1 \leftrightarrow Q$
    transition crosses over, lowering, in consequence,
    the population of the spin-up state. This results into a quench of the nonequilibrium excitation. 
    Due to the proximity to \set regime as compared to \Fig{fig:10}(a), a \coset feature (orange dotted line) appears 
    as a mirage of a spin-excitation.}
  \label{fig:12}
\end{figure}

We first investigate how \cot spectrum evolves
as we further
 \emph{approach} the \set regime from either side.
\Fig{fig:12}(a) shows the analogous of \Fig{fig:10}(a) 
but closer to the \set regime, at $V_{\g} = - 1.32$~V. 
A horizontal, $B$-field independent line appears (dotted green line
in the center-panel schematic) that terminates at $B_c \approx 4.5$~T,
precisely upon crossing the intra-triplet excitation (blue line).
This indicates that the excited triplet (spin $S=1$ perpendicular to the field, $M=0$)
lives long enough for a secondary \cot process to excite the system to the singlet state
(reducing the spin length to $S=0$).
Strong evidence for this is the termination of this line:
once the initial state ($M=0$ excited triplet) for this transition is no longer accessible for $B > B_c$,
the "nonequilibrium" cascade of transitions is interrupted.

We consistently observe this effect, also when approaching the \set regime from the side of the other charge state ($N=3$) with different spin.
In \Fig{fig:12}(b) we show the magnetic field spectrum taken at $V_{\g} = - 1.75$~V. 
Here the lowest $D_2$ excitation
gains strength\cite{foot:invisible} relative
 to \Fig{fig:10}(b).
In this case, the excited $D_1$ state is the starting point of a "nonequilibrium" cascade.
As for the previous case, it terminates when levels cross at $B \approx 4$~T
for similar reasons:
Once the $Q$ state gains occupation for $B > B_c$ (since the $D_1 \leftrightarrow Q$ transition becomes energetically more favorable)
the excited $M$-substates of the $D_1$ multiplet
are depleted causing the line to terminate.
In both charge states, the observed \cot current
gives an estimate for the spin-relaxation time,  $\tau_\text{rel} \gtrsim 10^{-9}$~s.

Nonequilibrium transitions can thus give rise to clear excitations at \emph{lower} energy
than expected from the simple selection-rule plus equilibrium arguments of \Sec{sec:spectro}.
In this type of processes, two \cot event ($\Gamma^2$) happen in sequence,
so that a total of four electrons are involved.\cite{foot:4electrons}
In this sense, the phenomena can be regarded as a single-molecule \emph{electronic pump-probe} experiment,
 that is,
the excess energy left behind by the first process (pump) allows the second process to reach states (probe) that 
would be otherwise inaccessible at the considered bias voltage.
This has been successfully applied in STM studies~\cite{Loth_NaturePhys.6/2010,Heinrich_Nat.Phys.12/2013}
for dynamical  spin-control.

\subsection{Mirages of spin transitions ``far from resonance''\label{sec:coset}}

\begin{figure*}[t]
  \includegraphics[width=0.99\textwidth]{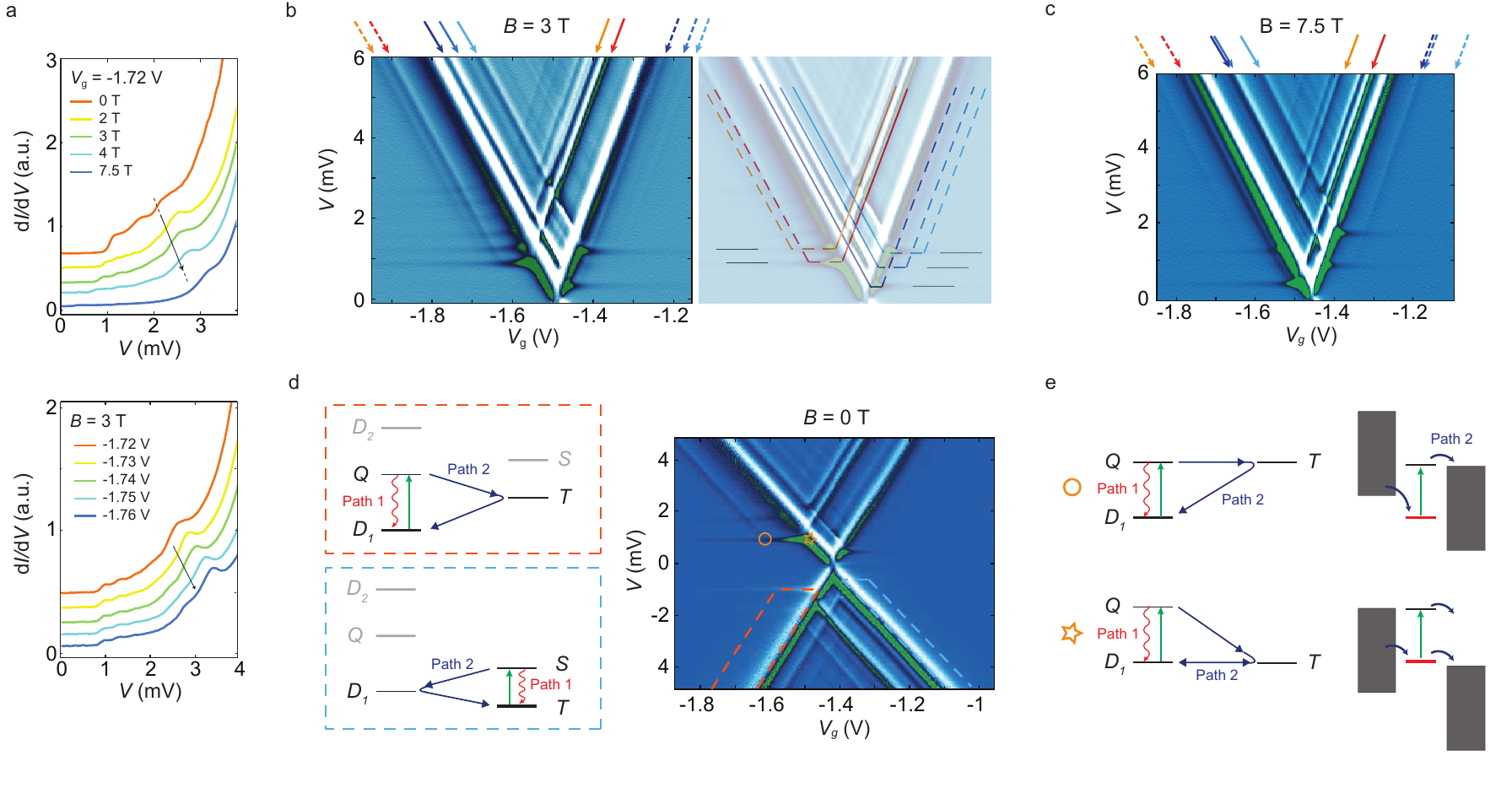}
  \caption{
    \textbf{Transport characteristics of spin \coset mirages}.
    (a)
    \didv spectra taken at $V_{\g} = - 1.72$~V for different $B$-fields 
    (upper) and at $B=3$~T for different values of $V_{\g}$ (lower).
    The step indicated by the black arrow Zeeman-splits as a regular magnetic transition.
    The same step moves to higher energies as a function of gate voltage
    and fixed magnetic field. (b) \didvv color map at $B = 3$~T. 
    Red (Blue) arrows and lines 
    indicate the \set excitations that extend their mirror images into the 		
    \cot regions of the $N$ ($N+1$) charge state.
    (c)
    \didvv color map at $B = 7.5$~T. 
    The $B$-field evolution of the mirror lines follows that of their real counterparts.
    (d)
    \didvv of the plot in \Fig{fig:10}(c).
    The cotunneling-assisted \set transport bands are highlighted.
    As illustrated in the schematics, mirror images are created when two or more relaxation paths compete.
    One path (red) involves the intra-molecular relaxation rate characteristic of the \cot regime.
    The second, alternative path (blue) requires a charge/spin fluctuation to the ($N\pm1$, $S_{N\pm1}$) state.
    (e)
    Energy and chemical potential schematics at the boundaries of the \coset bands.
    \set and \cot transport occur ($\medwhitestar$).
    More negative gate voltages shift the $T$ state higher in energy,
    forbidding \set and leaving only \coset and \cot competing.
    At the position indicated by $\medcircle$ further gating finally quenches \coset.
  }
  \label{fig:13}
\end{figure*}

\paragraph{Mirages.---}
We now further reduce the distance to resonance,
again coming from either side,
and \emph{enter} the crossover regime
discussed in \Sec{sec:crossover}.
We are, however, still ``well away from resonance'' by the linear-response  condition \eq{eq:naive}.

In the upper panel of \Fig{fig:13}(a) we show \didv traces taken at various magnetic fields
for a constant gate voltage \mbox{$V_{\g}=-1.72$~V}.
At high bias voltage the \didv steeply rises due to the onset of the main \set resonance.
Below this onset, we note a step-like excitation at $V= 2.1$~meV (black arrow)  
which shifts up in magnetic field with the same $g$-factor ($\approx 2$) as the other lower-lying \cot excitations.\cite{foot:partner}
If one adopts the \stm picture this excitation is attributed to the opening of an independent \cot ``channel''.
This attribution proves to be erroneous:
Keeping $B=3$~T fixed and varying the gate voltage (\Fig{fig:13}(a), lower panel),
we observe that the lower excitations are left unchanged,
whereas the higher one under consideration \emph{shifts linearly} with $V_{\g}$,
revealing that it is \emph{not} a \cot excitation.

This attribution to \cot
 can be further ruled out by looking at the full gate-voltage dependence in the stability diagram shown in the left panel of  \Fig{fig:13}(b).
The excitation (red arrow) has the same gate dependence as the \set resonances,
even though it is definitely not in the \qd regime by the linear-response criterion \eq{eq:naive}.
In fact, it is a \coset mirage of the \emph{same} lowest gate-voltage independent \cot excitation
as we explained in \Fig{fig:mirage}.
Its bias (energy) position does not provide information about the excitation energy $\Delta$:
depending on the energy level position
the mirage's excitation voltage $V^{*}$ can lie anywhere above the \cot threshold voltage $V=\Delta$,
see \Sec{sec:crossover}.

In the stability diagram in the right panel of \Fig{fig:13}(b), we connect
by dashed lines all the \coset resonances to their corresponding \set excitations according to the scheme in \Fig{fig:mirage}.
We find that mirages appear for virtually all spin-related excitations of the molecule.
The stability diagram in \Fig{fig:13}(c) [same color coding as in (b)]
shows that at high magnetic field $B=7.5$~T
these mirages persist. 

The clearly visible \coset resonances
mark the lines where 
the relaxation mechanism changes from ``virtual'' (\cot)
to ``real'' (\set) charging.
They indicate that any intrinsic relaxation is comparable or slower
than \set.\footnote{(If the intrinsic relaxation was much faster, compared to \set, it would dominate everywhere, giving a much smaller change in the current at \coset resonances.)}
Mirages are thus a signature of slow intramolecular relaxation,
in particular they indicate that the intrinsic relaxation time is bounded from below by
the magnitude of the observed \set currents
$\tau_\text{rel} \gtrsim \Gamma^{-1} \sim 10^{-11}$~s,
consistent with the sharper lower bound we obtained above from nonequilibrium \cot spectroscopy.

\paragraph{Spin relaxation.---}
To shed light on what the relaxation mechanism by transport entails in our device,
we return to the stability diagram for $B=0$~T,
which is shown as \didvv in the right panel of \Fig{fig:13}(d).
Highlighted at negative bias are the two crossover-regime bands within which \coset, rather than \cot, dominates the relaxation.
The left panel shows the different relaxation paths for these two bands.

Focusing on the orange band,
we start out on the far left of \Fig{fig:13}(d)
moving at fixed bias $V=1.2$~meV along the onset of inelastic \cot.
\Fig{fig:13}(e)
depicts the corresponding energies (left)
and energy differences (right).
Here, the molecule is in the spin-doublet $D_1$ ground state and is occasionally excited to the high-spin quartet $Q$ by \cot
from where it relaxes via path 1 ($10^{-9}$ s), again by \cot.
 
When reaching the circle ($\medcircle$) in \Fig{fig:13}(d)
the \emph{relaxation} mechanism changes:
path 1
is overridden by the faster 
relaxation path 2
($10^{-11}$ s)
which becomes energetically allowed [\Eq{eq:stableex}].
The top panel of \Fig{fig:13}(e) illustrates
that although the ground state $D_1$ is off-resonant (highlighted in red),
after exciting it by \cot to $Q$
---increasing the spin-length---
the system has enough \emph{spin-exchange} energy (green) to expel 
a single electron in a ``\emph{real}'' tunneling processes
leaving a \emph{charged} triplet state behind.

At the star ($\medwhitestar$) in \Fig{fig:13}(d)
the \emph{excitation} mechanism changes from \stm to \qd,
leaving the relaxation path unaltered.
Now the ground state $D_1$ becomes unstable with respect to ``real'' charging:
there is enough energy to expel an electron to the right electrode
and sequentially accept another one from the left.
We thus have an \qd \set transport cycle,
i.e., the stationary state is a statistical mixture of the $N$ and $N+1$ ground states.

\begin{figure}[H]
  \includegraphics[width=.99\columnwidth]{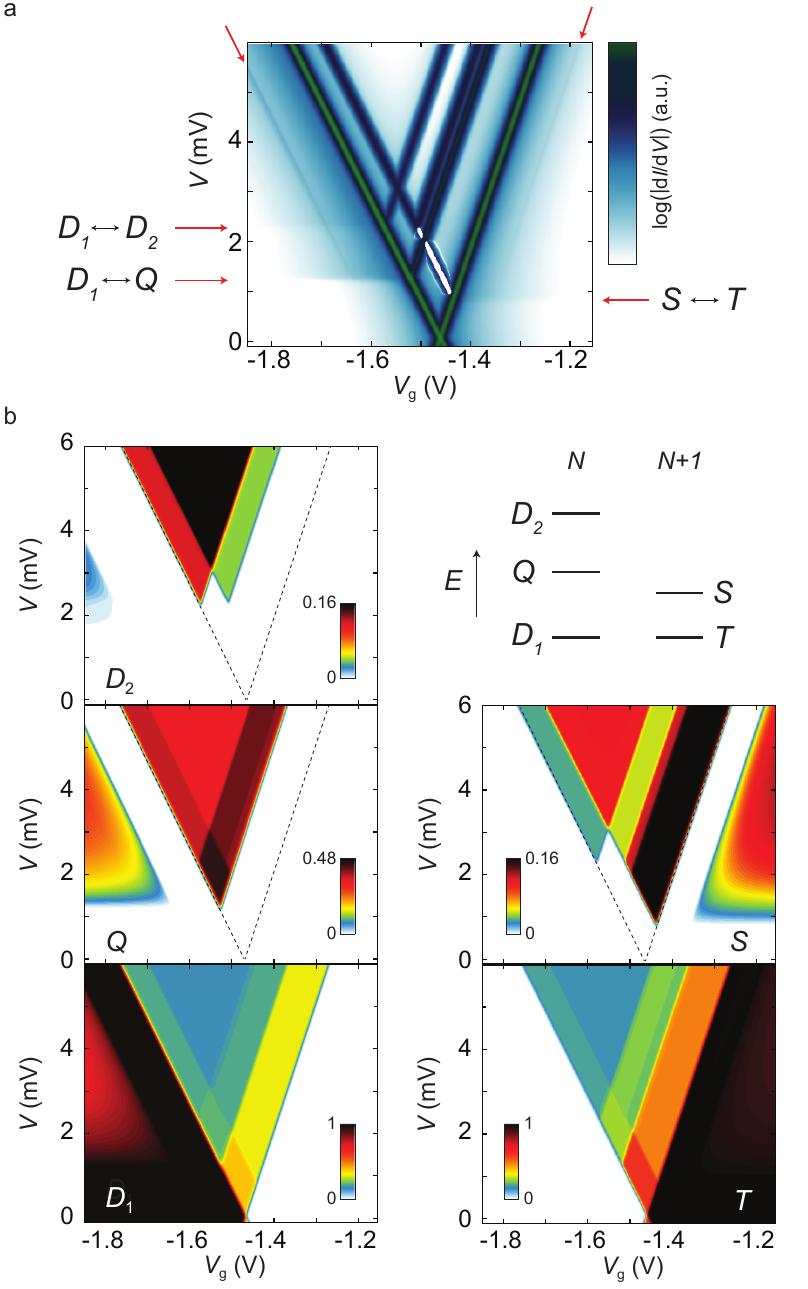}
  \caption{
    \textbf{Calculated stability diagram and multiplet occupations}.
    Result obtained from the full master equation \eq{eq:full}
    and its corresponding current formula 
    (not shown, see \Refs{Leijnse_Phys.Rev.B78/2008,Koller_Phys.Rev.B82/2010})
    for the same parameters as in \Fig{fig:11}(b).
    (a)
    Transport spectrum for $B=0$ corresponding to \Fig{fig:10}(c).
    (b)
    Corresponding color plots of the occupation probabilities of the five spin multiplets
    (probabilities summed of degenerate levels).
    The effective coupling $\Gamma^{*}$
    ---merely an overall scale factor in \Fig{fig:11}(b)---
    now controls the magnitude
    of the \cot and \coset current corrections \emph{relative} to the \set current.
    Although elaborate, these corrections still neglect nonperturbative broadening effects
    and must kept small for consistency by explicitly setting
    $\Gamma^{*} =  2.2 \cdot 10^{-3} \, T^{*} = 0.6 \text{ mK} = 5 \cdot 10^{-5}\text{ meV}$.
    %
    %
    More advanced master equation approaches
    based on renormalization-group\cite{foot:rg,Saptsov_Phys.Rev.B86/2012}  (RG) or hierarchical\cite{foot:hqme,Schinabeck16} (HQME) methods can deal with both this broadening and the corresponding larger currents.
  }
  \label{fig:14}
\end{figure}

The \coset regime is delimited by mirage resonances and situated
between the two positions $\medcircle$ and $\medwhitestar$. 
Failure to identify the difference between this "band" and the pure \cot happening on the left of $\medcircle$,
besides yielding a wrong qualitative spin multiplet structure, leads to an overestimation of the relaxation time:
in the \coset regime the spin-excitations created by inelastic \cot are \emph{quenched}.

\paragraph{Quenching of spin-excitations.---}

We now assess this quenching in detail for the experimental situation
by a calculation based on the  master equation~\cite{Leijnse_Phys.Rev.B78/2008,Koller_Phys.Rev.B82/2010} \eq{eq:full} that includes all $\Gamma$ and $\Gamma^2$ processes
using the model determined earlier [\Eqs{eq:energies_3}-\eq{eq:rates_S}],
simulating the broadening as before by an effective temperature [\Fig{fig:11}].

The computed conductance for $B=0$ is shown in \Fig{fig:14}(a).
Besides the \set excitations ---including the NDC effect--- obtained earlier in \Fig{fig:11}(c),
we capture the main features of the experimental data in \Fig{fig:11}(c) and \Fig{fig:13}(d):
the three horizontal \cot excitations and two prominent \coset lines.

We can now explore the nonequilibrium occupations of the five spin-multiplets
 as the transport spectrum is traversed.
These are shown in \Fig{fig:14}(b).
The lowest panels show in
the left (right) \stm regime the ground multiplet $D_1$ with $N$ electrons
($T$ with $N+1$ electrons) is occupied with probability 1 at low bias voltage (black regions).
In contrast, in the \qd regime these two ground states are both partially occupied due to \set processes.
We compare the occupations  along three different vertical \didv line cuts  in \Fig{fig:14}(a).

(i)
Increasing the bias voltage in the \qd regime,
starting from $V_{\g}=-1.46$~V,
one first encounters in \Fig{fig:14}(a) a \didv dip (NDC, white).
This is caused by the occupation of the $S$ state,
as the $S$-panel in \Fig{fig:14}(b) shows.
This drains so much probability from the $T$ multiplet [with a higher transition rate to the $D_1$ multiplet,
\Eq{eq:rates_T}-\eq{eq:rates_S}] that the current goes down.
Increasing the bias further depopulates the $S$ state again,
thereby restoring the \set current through a series of \didv peaks.

(ii)
Increasing the bias voltage starting from the right \stm regime
the excited $S$-state becomes populated by \cot
decreasing the average spin-length of the molecule.
When crossing the \coset resonance at higher bias this excitation
is \emph{completely quenched} (white diagonal band)
 well before reaching the \qd regime,
\emph{enhances} the molecular spin, restoring the triplet.

(iii)
When starting from the left \stm regime,
the population of the excited $Q$-state {enhances} the average spin-length of the molecule.
As before, crossing the \coset resonance at higher bias \emph{quenches} this excitation.
Now this \emph{reduces} molecular spin, restoring the doublet.
Along the way, the $D_2$ state also becomes occupied by \cot and subsequently quenched by \coset.
Because of its higher energy, the white \coset band in the $D_2$-panel of \Fig{fig:14}(b) is much broader.

\begin{figure}[H]
  \includegraphics[width=0.99\columnwidth]{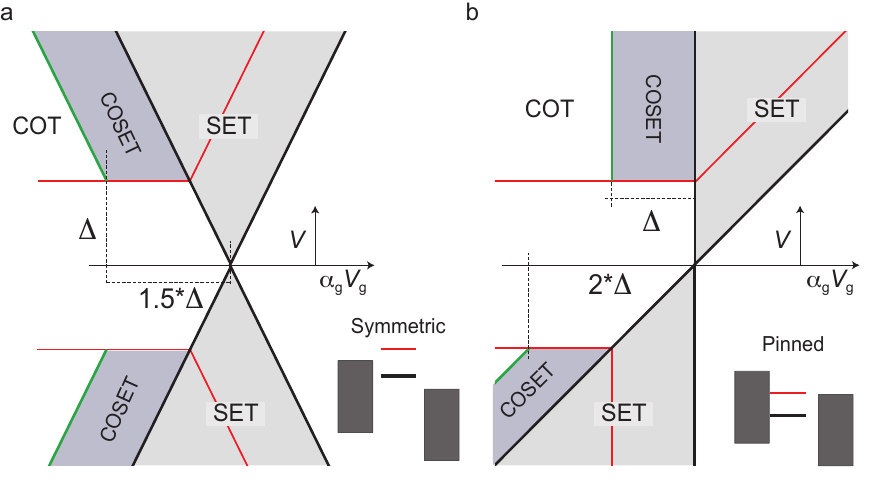}
  \caption{
    \textbf{How far is ``off-resonant'' ?}
    (a)
    Energy-energy stability diagram
    for symmetric capacitive coupling
    ($C_{\L}=C_{\R} \gg C_{\g}$)
    characteristic of molecular QD devices.
    (b)
    Strongly asymmetric couplings,
    ($C_{\L} \gg C_{\R}$).
    This is typical for molecular STM junctions,
    where the energy levels ``pin'' to one electrode (substrate),
    leaving the tip electrode to act as a probe.
    In this case the energy $\alpha_{\g} V_{\g}$ represents
    the \emph{level alignment} with the Fermi-energy.
  }
  \label{fig:15}
\end{figure}

\subsection{How far is ``off-resonant'' ?}

The results show that the widths of the two bands
 where the \cot excitations are quenched by \coset
are unrelated to the width of the \set resonances,
set by the maximum of $\Gamma$ and $T$.
They are, instead, set by the \emph{excitation spectrum} one wishes to probe.

In \Fig{fig:15}(a) we quantify how far the energy level has to be detuned from resonance
in order avoid this quenching in our molecular QD device structure.
When this detuning lies in the window $- 1.5 \Delta < \alpha_{\g} V_{\g} < - 0.5 \Delta$
one is sure to run into the \coset band with increasing bias.
Only for $-\alpha_{\g} V_{\g} < 1.5 \Delta$ there is a finite window where the excitation is not quenched.
For the excitations $T$, $Q$, $D_2$ in our experiment, this amounts to
$ $, $ $, and $ $ times the \set resonance width.

In \Fig{fig:15}(b) we show the corresponding construction for strong capacitive asymmetry
typical of STM setups. 
To avoid quenching for any bias polarity, one now needs to stay further away from resonance
$-\alpha_{\g} V_{\g} < 2 \Delta$.
Interestingly,
for $-2 \Delta < \alpha_{\g} V_{\g} < - \Delta$
\cot excitation at forward $V=\Delta$ is not quenched,
whereas at reverse bias $V=-\Delta$ it is.
For asymmetric junctions, the \coset mechanism thus leads to a \emph{strong bias-polarity dependence} of relaxation of excitations in the nominal \stm regime.
For $- \Delta < \alpha_{\g} V_{\g}$ one is sure to run into the \coset band for forward bias.

Whereas in the present experiment we encountered relatively low-lying spin-excitations ($\Delta \sim $~few meV)
atomic and molecular devices can boast such excitations up to tens of meV.
To gauge the impact of \coset mirage resonances, consider an excitation at $\Delta \sim 25$~meV
that we wish to populate by \cot, e.g., for the purpose of spin-pumping~\cite{Loth_NaturePhys.6/2010,Heinrich_Nat.Phys.12/2013}.
To avoid the quenching of this excitation $V=\Delta$ the distance to the Fermi-energy at $V=0$ (level-alignment) needs to exceed \emph{room temperature},
even when operating the device at mK temperatures.
For vibrational and electronic excitations on the 100~meV scale the implications are more severe.
Moreover, even for excitations that do satisfy these constraints, cascades of "nonequilibrium" \cot excitation may ---if even higher excitations are available (e.g., vibrations)--- provide a path to excitations that do decay by \set processes.
Whereas all these effects can be phrased loosely as ``heating'' in this paper we demonstrated the discrete nature of these processes, their \emph{in-situ} tuneability, and the role they play as a spectroscopic tool.

\section{Summary and Outlook\label{sec:discussion}}

We have used electron transport on a single-molecule system to comprehensively characterize 
the spin degree of freedom and its interaction with the tunneling electrons.  
Three key points ---applicable to a large class of systems--- emerged with particular prominence:

(i) Combining \set and \cot spectroscopy in a single stable device
provides new tools for determining spin properties \emph{within} and \emph{across} molecular redox states.
This is crucially relevant for the understanding of the different spin-relaxation mechanisms,
even in a \emph{single} redox state.

(ii) Nonequilibrium pump-probe electron excitation using two \cot processes (four electrons)
was demonstrated in our three-terminal molecular device
and signals a substantial intrinsic spin relaxation time of about 1 ns,
much larger than the transport times.

(iii) Mirages of resonances arise from the nontrivial interplay of \set and \cot.
These \coset resonances signal a sharp increase of the relaxation rate and can occur far away from resonance (many times the resonance width).
This limits the regime where spin-pumping works by quenching nonequilibrium populations created by a \cot current.

The appearance of a mirage of a certain \cot excitation indicates
that the relaxation of the corresponding molecular degree of freedom dominates over
all possible unwanted, intrinsic mechanism.
Thus, ``good devices show mirages'' and ``even better devices'' show nonequilibrium \cot transitions.

Energy level control turns out to be essential for ``imaging'' in energy space,
distinguishing mirages from real excitations.
Whereas real-space imaging  seems to be of little help in this respect,
the mechanical gating possible with scanning probes overcomes this problem.
However, even when energy-level control is available,
spectroscopy of molecular junctions still requires extreme care
as we illustrated in \Sec{sec:break} by several examples that break spectroscopic rules.
Moreover, our work underlines that level alignment has to be treated on a more similar footing as 
as coupling ($\Gamma$) and temperature ($T$) broadening
in the engineering of molecular spin structures and their spin-relaxation rates~\cite{Loth_NaturePhys.6/2010, Heinrich_Nat.Phys.12/2013}.

Beyond electron charge transport, recent theoretical work~\cite{Gergs_Phys.Rev.B91/2015,Gergs17a}
has pointed out that importance of \coset is amplified when moving to nanoscale transport of \emph{heat}~\cite{Galperin_J.Phys.:Condens.Matter19/2007}.
Whereas in charge transport all electrons carry the same charge,
in energy transport electrons involved in \coset processes effectively can carry a quite different energy
from that acquired in a  \cot process only and therefore dominate energy currents~\cite{Gergs_Phys.Rev.B91/2015,Gergs17a}.
Thus, the sensitivity to spin-relaxation processes is dramatically increased in heat transport,
indicating an interesting avenue~\cite{Gergs17a} for a \emph{spin-caloritronics}~\cite{Bauer_NatureMater.11/2012} on the nanoscale.

\acknowledgments
We thank A. Cornia for the synthesis of the molecules,
M. Leijnse and M. Josefsson for assistance with the calculations,
and S. Lounis, M. dos Santos Dias and T. Esat for discussions.
We acknowledge financial support by the Dutch Organization for Fundamental research (NWO/FOM)
and an advanced ERC grant (Mols@Mols).
M. M. acknowledges financial support from the Polish Ministry of Science and Higher Education through a young scientist fellowship (0066/E-336/9/2014),
and from the Polish Ministry of Science and Education as Iuventus Plus project (IP2014 030973) in years 2015-2017.
E.B. thanks funds from the EU FP7 program, Project 618082 ACMOL through a NWO-VENI fellowship.


\begin{thebibliography}{265}
\bibitem [{foo({\natexlab{a}})}]{foot:rates}%
  \BibitemOpen
  \bibinfo {note} {These rates can be brought into a form similar to those
  obtained in the \asw and $\mathcal{T}$-matrix approach, allowing some of
  advantages of the latter to be exploited.}\BibitemShut {Stop}%
\bibitem [{foo({\natexlab{c}})}]{foot:physisorption}%
  \BibitemOpen
  \bibinfo {note} {Physisorption is the dominant binding mechanism for the
  molecular species used here, given no chemically-active linking groups are
  present. Chemisorption is the other common alternative used in molecular
  electronics where a chemical bond forms between the molecule and the
  electrode.}\BibitemShut {Stop}%
\bibitem [{foo({\natexlab{d}})}]{foot:greenline}%
  \BibitemOpen
  \bibinfo {note} {The dashed green line in (a) can also be included by
  allowing a slight breaking of spin-conservation, enabling the $S
  \leftrightarrow Q$ transition. This more detailed modeling goes beyond the
  level of detail required for the \coset and \cot features on which we
  focus.}\BibitemShut {Stop}%
\bibitem [{foo({\natexlab{e}})}]{foot:invisible}%
  \BibitemOpen
  \bibinfo {note} {{Since this enhancement of the excited $D_1$ state is small
  compared to that of the other "equilibrium" excitations it is not revealed by
  \Fig{fig:10}(b).}}\BibitemShut {Stop}%
\bibitem [{foo({\natexlab{f}})}]{foot:4electrons}%
  \BibitemOpen
  \bibinfo {note} {Note that this is not a single $\Gamma^4$ processes, since
  the two $\Gamma^2$ steps occur in sequence. This is similar to the \coset
  which is not a $\Gamma^3$ process, but a sequence of two processes, one of
  the order of $\Gamma^2$ and the other of $\Gamma$.}\BibitemShut {Stop}%
\bibitem [{foo({\natexlab{g}})}]{foot:partner}%
  \BibitemOpen
  \bibinfo {note} {The downshifting partner excitation is visible at a
  different gate voltage in \Fig{fig:12}(b) (labeled as \coset) and in the
  lower panel of \Fig{fig:13}(a).}\BibitemShut {Stop}%
\bibitem [{foo({\natexlab{h}})}]{foot:rg}%
  \BibitemOpen
  \bibinfo {note} {In a renormalization group (RG) treatment of the master
  equation the broadening arises naturally. First, in uniform way as the sum of
  all tunnel rates in the discrete RG approximation. Then during a continuous
  RG flow the broadening is corrected in an energy dependent way.}\BibitemShut
  {Stop}%
\bibitem [{foo({\natexlab{i}})}]{foot:hqme}%
  \BibitemOpen
  \bibinfo {note} {The nonperturbative hierarchical density-operator approach
  (HQME) can also deal with the broadening, as demonstrated both for \set and
  \coset resonances in the cited reference.}\BibitemShut {Stop}%
\end{thebibliography}

%

\end{document}